\documentclass[12pt]{elsarticle}

\usepackage{lineno,hyperref}
\modulolinenumbers[5]
\usepackage{graphicx}
\usepackage{amssymb}
\usepackage{epstopdf}
\usepackage{amsmath}
\usepackage{bm}
\usepackage{float}

\journal{Nucl. Phys. A}

\begin{document}

\begin{frontmatter}

\title{Application of the Thomas Fermi Quark Model to Multiquark Mesons}

\author[mymainaddress,mysecondadd,mythirdadd]{Suman Baral}
\address[mymainaddress]{Department of Physics, Baylor University, Waco, TX USA 76798-7316}
\address[mysecondadd]{Everest Institute of Science and Technology, 343 Ranibari marga, Kathmandu, Nepal}
\address[mythirdadd]{Neural Innovations LLC, 3225 Skinner Dr., Lorena, TX USA 76655}
\ead{suman@everestsciencetech.com; suman@neuralinnovations.io}

\author[mymainaddress]{Walter Wilcox\corref{mycorrespondingauthor}}
\ead{walter\_wilcox@baylor.edu}
\cortext[mycorrespondingauthor]{Corresponding author}

\begin{abstract}
The possibility of the existence of mesons with two or more quark-antiquark pairs is investigated with a new application of the Thomas-Fermi (TF) statistical quark model. Quark color couplings are treated in a mean field manner similar to a previous application to baryons, and short and concise expressions for energies are derived. We find that, on average, quarks only interact with antiquarks in such systems. The TF differential equation is constructed and systems with heavy-light quark content are examined. Three types of mesonic systems are defined. In the case of charm quarks, multi-charmonium, multi-$Z$ meson and multi-$D$ meson family types are examined. System analogs for bottom quarks are also constructed. Quantitative trends for system energies of mesonic quark matter are extracted as a function of the number of quark pairs. We find indications from energy plots that multi-$Z$ type mesons (and their bottom quark analogs) are actually stable for a range of quark number pairs. At this initial stage we have not yet included explicit spin interaction couplings between quarks, but we can take one level of degeneracy into account in our two-inequivalent TF function construction. 
\end{abstract}

\begin{keyword}
quark model \sep Thomas-Fermi \sep tetraquarks \sep quark matter
\PACS 12.39.-x\sep 12.39.Mk\sep 12.40.Ee \sep 21.65.Qr\\
{\it Preprint:} BU-HEPP-19-04
\end{keyword}

\end{frontmatter}


\section{Introduction}

Lattice QCD is a very important tool used by particle physicists to investigate the properties of baryons and mesons. Lattice techniques\cite{Nilmani,N2,N3,N4,N5,N6} are presently being employed to understand and elucidate the pentaquark and tetraquark observations made by Belle\cite{Belle0,Belle,B2,B3}, BESIII\cite{BESIII,BES2}, LHCb\cite{LHCb00,LHCb,LHCb1,LHCb0,LHC2}, and other collaborations. However, as the quark content increases, it becomes computationally expensive and time-intensive to do the lattice calculations. Every state must be investigated separately, which means a great deal of analysis on Wick contractions and specialized computer coding. In addition, as one adds more quarks, the states will become larger and the lattice used must also increase in volume. There is therefore a need for reliable quark models that can give an overview of many states to help guide these expensive lattice calculations. The MIT bag model\cite{Mit,Mads1,Mads2}, the Nambu-Jona-Lasinio\cite{Njl} model and explicit tetraquark potential models\cite{Maiani} are some of these. Another approach, the Thomas-Fermi (TF) statistical model has been amazingly successful in the explanation of atomic spectra and structure, as well as nuclear applications. Our group has adopted the TF model and applied it to collections of many quarks\cite{WW1,WW2,Baral,singapore}. One advantage our model has over bag models is the inclusion of nonperturbative Coulombic interactions. One would expect that the TF quark model would become increasingly accurate as the number of constituents is increased, as a statistical treatment is more justified. The main usefulness will be to detect systematic trends as the parameters of the model are varied. It could also be key to identifying {\it families} of bound states, rather than individual cases. The TF quark model has already been used to investigate multi-quark states of baryons\cite{WW2}. In this paper we have extended the TF quark model to mesonic states in order to investigate the stability of families built from some existing mesons and observed new exotic states, concentrating on heavy-light quark combinations. Although our model is nonrelativistic, we will see that this assumption is actually numerically consistent as quark content is increased.

Our paper is organized as follows. In Sec.~\ref{sec1} we will define the energies of the model in terms of the TF density of states. In Sec.~\ref{sec2} we will examine the classical color couplings in mean field theory and systematically determine the probabilities of interactions for quark-quark, quark-antiquark and antiquark-antiquark interactions. We obtain the system energies in Sec.~\ref{sec3}, the TF quark equations in Sec.~\ref{sec4}, and re-characterize the model in terms of new dimensionless variables in Sec.~\ref{sec5}. The application of the model to heavy quark-light quark mesonic matter is initiated in Sec.~\ref{sec6}, where we define three types of multi quark-pair families involving charm quarks: charmonium (\lq\lq Case 1"), $Z$-meson type (\lq\lq Case 2") and $D$-meson type (\lq\lq Case 3"). These states as well as their bottom-quark analogs are constructed in Sec.~\ref{sec6}, where we examine the energy slopes to determine family stability. Numerical results and discussions are presented in Secs.~\ref{sec7} and \ref{sec8}, and concluding thoughts and remarks are presented in Sec.~\ref{sec9}.

\section{The TF meson model} \label{sec1}

The TF statistical model is a semiclassical quantum mechanical theory developed for many-fermion systems. In this model the assumption is made that quarks are distributed uniformly in each volume element $\Delta V$, while at the same time the quark density $n_q(r)$ can vary from one small volume element to the next. For a small volume element $\Delta V$, and for the system of quarks in it's ground state, we can fill out a spherical momentum space volume $V_F$ up to the Fermi momentum $p_F$, and thus
\begin{equation}\label{ab694}
\begin{aligned}
V_{F} \equiv \frac{4}{3}\pi p_{F}^{3}\left( \vec{r}\right).
\end{aligned}
\end{equation}
The corresponding phase space volume is,
\begin{equation}
\begin{aligned}
\Delta V_{ph}  \equiv V_{F}\Delta V = \frac{4}{3}\pi p_{F}^{3}{(\vec r)}\Delta V.
 \end{aligned}
\end{equation}
Let's say $g^I$ is the degeneracy of a quark flavor $I$. Then quarks in $\Delta V_{{p}{h}}$ are distributed uniformly with $g^I$ quarks per $h^3$ of this phase space volume, where $h$ is Planck's constant. The number of quarks in $\Delta V_{{p}{h}}$ is
\begin{equation}\label{ab709}
\begin{aligned}
\Delta N \equiv \frac{g^{I}}{h^3}\Delta V_{p{h}}  = \frac{4\pi g^{I}}{3{h^3}} p_{F}^{3}{(\vec r)}\Delta V.
\end{aligned}
\end{equation}
The number of quarks in $\Delta V$ is
\begin{equation} \label{kali}
\begin{aligned}
\Delta N \equiv n_q{(\vec r)}\Delta V,
\end{aligned}
\end{equation}
where
 $n_q{(\vec r)} $ is the quark density.
Equating Eqs.~(\ref{ab709}) and (\ref{kali}), we obtain
\begin{equation}\label{hero}
\begin{aligned}
n_q{(\vec r)}=\frac{4\pi g^{I}}{3{h^3}}{p_{F}^{3}(\vec r)}.
\end{aligned}
\end{equation}
The fraction of quarks at position $\vec{r}$ that have momentum between $p$ and $p+dp$ in spherical momentum space is
 \begin{equation}
 F_{(\vec r)}(p)dp =
\begin{cases}
      \frac{4\pi p^2 dp}{\frac{4}{3}\pi p_{F}^{3}(\vec r)}, \;\;\;\;\;\;\;\;\text{if}\;p \leq p_{F}(\vec r), \\
      0, \;\;\;\;\;\;\;\;\;\;\;\;\;\;\;\;\;\;\text{otherwise}.
     \end{cases}
\end{equation}
Using the classical expression for kinetic energy of a quark with mass $m_I$, the kinetic energy per unit volume for a system of quarks is
\begin{equation}\label{E7}
\begin{aligned}
 t(\vec{r})  = \int^{p_F} \frac{p^2}{2m_I}n_q(\vec{r}) F_{(\vec r)}(p)dp  = C_F \left( n_q(\vec{r})\right)^{5/3},
\end{aligned}
\end{equation}
where
\begin{equation}
C_F = \frac{\left( 6\pi^2 \hbar^3 \right)^{5/3}}{20 \pi^2 \hbar^3 m_I\left( g^{I}\right)^{2/3}}.
\end{equation}
Eq.~(\ref{E7}) shows that kinetic energy per volume is proportional to the $5/3$-rd power of quark density. To obtain the total kinetic energy ($T$), we will have to integrate this expression over all the spatial volume:
 \begin{equation}\label{abc214}
\begin{aligned}
T = \int  t(\vec{r}) \,d^3r.
\end{aligned}
\end{equation}
On the other hand, the total potential energy ($U$) due to the coulombic interactions of system of quarks with one another is given by
\begin{equation}
\begin{aligned}
U= V_N \int \int \frac { n_q(\vec{r}) n_q(\vec{r}\,') }  {|\vec{r}-\vec{r}\,'| }d^3r \,d^3r',
\end{aligned}
\end{equation}
where $V_N$ is a factor depending on the type of  interaction between quarks. It will be explained in the sections to follow. The total energy for a system of quarks is therefore
\begin{equation}
\begin{aligned}
E &= T+U \\ &
= C_F \int \left(n_q(\vec{r}) \right)^{5/3} d^3r + V_N \int \int \frac { n_q(\vec{r}) n_q(\vec{r}\,') }  {|\vec{r}-\vec{r}\,'| }d^3r d^3r'.
\end{aligned}
 \end{equation}
In order to minimize energy while keeping the number of quarks constant, we can add a Lagrange multiplier term of the form $\lambda \left( \int n_q(\vec{r}) d^3(r) - N_q\right)$ to the expression for total energy, where $N_q$ is the fixed total number of quarks in the given system.

A given quark actually has a definite color and flavor. In the sections to follow we will replace the number density of quarks, $n_q(\vec{r})$, by the quantity ${n_i^{I}(\vec{r})}$, where color is represented by index $i$ and flavor by index $I$. It will then be summed over flavor and color indices to account for the energies of all of the quarks in the given system.

\section{Residual Coulombic coupling and Interaction probabilities}\label{sec2}

The color couplings of quarks and antiquarks in our model originate from the Coulombic interactions expected at the classical level \cite{Sikivie}. In the following we define $\eta$ to be the number of quark/antiquark pairs in the meson, which is assumed to be a color singlet. In addition, $g$ represents the strong coupling constant.

The types of interactions between the particles can then be categorized as:\\

\textbf{Color-Color Repulsion (CCR)}

Interactions between quarks with same colors is repulsive with coupling constant $4/3g^2$. The interactions are red-red ($rr$), green-green ($gg$) and blue-blue ($bb$).\\

\textbf{Color-Color Attraction (CCA)}

Interactions between quarks with different colors is attractive with coupling constant $-2/3g^2$. The interactions are $rb$, $rg$, $bg$, $br$, $gr$,  and  $gb$.\\

\textbf{Color-Anticolor Repulsion (CAR)}

Interactions between quarks and antiquarks with different color/anticolors is repulsive with coupling constant $2/3g^2$. The interactions are $r\bar{b}$, $r\bar{g}$, $b\bar{g}$, $b\bar{r}$, $g\bar{r}$, and $g\bar{b}$.\\

\textbf{Color-Anticolor Attraction (CAA)}

Interactions between quarks and antiquarks with same color/anticolors is attractive with coupling constant $-4/3g^2$. The interactions are $r\bar{r}$, $b\bar{b}$ and $g\bar{g}$.\\

\textbf{Anticolor-Anticolor Repulsion (AAR)}

Interactions between antiquarks with same anticolors is repulsive with coupling constant $4/3g^2$. The interactions are $\bar{r}\bar{r}$, $\bar{b}\bar{b}$  and  $\bar{g}\bar{g}$.\\

\textbf{Anticolor-Anticolor Attraction (AAA)}

Interactions between antiquarks with different anticolors is attractive with coupling constant $-2/3g^2$. The interactions are $\bar{r}\bar{b}$, $\bar{r}\bar{g}$, $\bar{b}\bar{g}$, $\bar{b}\bar{r}$, $\bar{g}\bar{r}$, and $\bar{g}\bar{b}$.\\

When $\eta$ pairs of quarks interact with each other, the six types of color interactions appear with different probabilities. So, we need to find a way to count and formulate color-interaction probabilities in order to apply the TF model for mesonic matter.

Each of the $\eta$ pairs must be a color singlet and each singlet can be achieved three different ways: red-antired, green-antigreen and blue-antiblue. We will term each quark-antiquark pair a \lq\lq pocket". Out of $\eta$ pockets let us say we have $x$ number of red-antired, $y$ number of blue-antiblue and $z= \eta -x -y$ number of green-antigreen combinations, which will be represented by $R$, $B$ and $G$, respectively. This is depicted in Table~\ref{Table2}.

\begin{table}[ht!]
\centering
\begin{tabular}{ |c|c|c|}
\hline
Number of pockets & Type of pocket  &  Representation\\
\hline
 $x$ & $r\bar{r}$ &  $ R$  \\
\hline
 $y$ & $b\bar{b} $ &  $ B$\\
 \hline
 $z$ & $g\bar{g}$ &  $G$\\
\hline
  \end{tabular}
 \caption{Color singlet counting and representation.}
\label{Table2}
\end{table}

When two $R$ pockets interact for example, there are four color combinations as depicted in the left box of Table~\ref{Table358}. Out of these, $r\bar{r}$ and $\bar{r}r$ belong to interaction type {CAA}, $rr$ to CCR and $\bar{r}\bar{r}$ to AAR. The middle box $RB$ depicts color combinations when different types of pocket interact. The rightmost box $R$ represents a color combination within this pocket, which means there is interaction type CAA only. These color combinations are counted and depicted in Table~\ref{Table327}. Of course $BB$ and $GG$ have the same interaction numbers as in the left box. In addition, $BG$ and $GR$ have same interaction numbers as in the middle box.

\begin{table}[ht!]
\centering
\begin{tabular}{ |c|c|c|}
\hline
$\bf{RR}$ & $\bf{r}$  &  $\bf{\bar{r}}$\\
\hline
 $\bf{r}$ & $rr$  &  $ r\bar{r}$\\
 \hline
 $\bf{\bar{r}}$ & $\bar{r}r$ &  $\bar{r}\bar{r} $\\
\hline
  \end{tabular}
\begin{tabular}{ |c|c|c|}
\hline
$\bf{RB}$ & $\bf{r}$  &  $\bf{\bar{r}}$\\
\hline
 $\bf{b}$ & $br$  &  $ b\bar{r}$\\
 \hline
 $\bf{\bar{b}}$ & $\bar{b}r$ &  $\bar{b}\bar{r} $\\
\hline
  \end{tabular}
\begin{tabular}{ |c|c|}
\hline
$\bf{R}$ & $\bf{r}$  \\
\hline
 $\bf{\bar{r}}$ & $\bar{r}r$ \\
\hline
  \end{tabular}
 \caption{Color interactions for pockets of the same color, pockets of different color and within a pocket.}
\label{Table358}
\end{table}

In Table~\ref{Table327} the six interaction types are given an index. The last three columns give the number of interactions contained in the previous table. 
\begin{table}[ht!]
  \begin{tabular}{ |c|c|c|c|c|}
\hline
Index(i) & Interaction type & $(RR)_i$ pocket & $(RB)_i$ pocket & $(R)_i$ pocket\\
\hline
1 & CCR & 1 &0 &0 \\
\hline
2 & CCA &  0&1&0\\
 \hline
3 & CAR &  0&2&0\\
\hline
4 & CAA &   2 &0&1\\
\hline
5 & AAR &  1&0&0\\
 \hline
6 & AAA & 0 &1&0\\
\hline
  \end{tabular}
 \caption{Color interaction counting and indices.}
\label{Table327}
\end{table}

There are $x(x-1)/2$ number of ways the $RR$ type of interaction takes place. Similarly, ${y(y-1)}/{2}$ and ${z(z-1)}/{2}$ are the number of times $BB$ and $GG$ happens. Also, $RB$, $BG$ and $RG$ type of interactions occur $xy$, $yz$ and $xz$ times respectively.

In the next step we varied $x$ from 0 to $\eta$ and $y$ from 0 to $\eta -x$ thereby giving equal footing to all the color combinations and counted all the possible color interactions, $E_i$. The corresponding expression is given in the equation below.
\begin{equation}\label{ab900}
\begin{aligned}
&E_i = \sum_{x=0}^{\eta} \sum_{y=0}^{\eta-x} \frac{\eta!}{x! y! z!}\times \\ &  \left \{   \frac{ x(x-1)+y(y-1)+z(z-1)}{2} {(RR)}_i + \left( xy + yz + zx \right) {(RB)}_i + \eta {(R)}_i \right \} .
    \end{aligned}
\end{equation}
Here $E_i$ refers to total occurrence of the $i^{th}$ color coupling. For example, $E_1$ refers to total number of times the AAA type of interaction occurs out of $3^{\eta}\times \eta (2\eta -1)$ possibilities.

To understand this equation, let us break it into two parts. First, if we take the terms inside the curly braces and add over the indices, we get $\eta(2\eta-1)$, i.e.,
\begin{equation}
\begin{aligned}
 \sum_i &\left \{   \frac{ x(x-1)+y(y-1)+z(z-1)}{2} {(RR)}_i + \left( xy + yz + zx \right) {(RB)}_i + \eta {(R)}_i \right \}\\
 &= \eta\left( 2\eta -1 \right),
 \end{aligned}
\end{equation}
regardless of $x$, $y$ and $z$ values as long as the constraint $x+y+z= \eta$ is satisfied. This is just the number of ways $2\eta$ quarks can be paired. Second, if we replace the curly braces by unity, we get $3^\eta$, the total number of pocket combinations in a meson with $\eta$ pairs of quarks. Thus, $E_i$ gives the total occurrence of the $i^{th}$ color interaction out of $3^\eta\eta(2\eta-1)$ possibilities. 

\begin{table}[ht!]
\centering
\begin{tabular}{ |c|c||c|c||c|c|}
\hline
 $(C)_i$ &Couplings & $(E)_i$ & Outcomes &  $(P)_i$ & Probabilities \\
\hline
 $C_1$&$\frac{4}{3}g^2$ &  $E_1$ & $3^{\eta -1} \frac{\eta (\eta-1)}{2}$ & $P_1$& $ \frac{\eta -1}{6(2\eta -1)}$\\
 \hline
$C_2$&$-\frac{2}{3}g^2$  &  $E_2$ & $3^{\eta -1} {\eta (\eta-1)}{}$ & $P_2$& $ \frac{\eta -1}{3(2\eta -1)}$\\
 \hline
$C_3$&$\frac{2}{3}g^2$ & $E_3$ & $3^{\eta -1} {2\eta (\eta-1)}{}$ &  $P_3$& $ \frac{2(\eta -1)}{3(2\eta -1)}$\\
 \hline
 $C_4$&$-\frac{4}{3}g^2$  &  $E_4$ & $3^{\eta -1} {\eta (\eta+2)}{}$ & $P_4$&$ \frac{\eta +2}{3(2\eta -1)}$ \\
 \hline
 $C_5$&$-\frac{4}{3}g^2$  & $E_5$ & $3^{\eta -1} \frac{\eta (\eta-1)}{2}$ & $P_5$&$ \frac{\eta -1}{6(2\eta -1)}$\\
 \hline
 $C_6$&$-\frac{2}{3}g^2$  & $E_6$ & $3^{\eta -1} {\eta (\eta-1)}{}$  & $P_6$&$ \frac{\eta -1}{3(2\eta -1)}$ \\
\hline
 \end{tabular}
 \caption{The coupling strength ($C_i$), total outcome ($E_i$) and probabilities ($P_i$) for all six interactions.}
\label{Table4}
\end{table}

After solving Eq.~(\ref{ab900}) using {\it Mathematica}, we obtained the outcomes, $E_i$. They are listed in Table~\ref{Table4}. In the same table, we divided each event by $3^{\eta} \eta (2\eta -1)$ and obtained probabilities for each type of interaction. In addition, the coupling strength, $C_i$, is defined for each interaction type.

If we add the product of coupling and probabilities from Table~\ref{Table4}, we get $-{\frac{4}{3}g^2}/{(2\eta -1)}$, very similar to the baryon case\cite{WW2}. The negative sign indicates that the system is attractive because of the collective residual color coupling alone, even in the absence of volume pressure. This gives rise to a type of matter that is bound, but does not correspond to confined mesonic matter, as discussed in Ref.~\cite{WW1}. We are interested in confined matter and will need to add a pressure term to the energy to enforce this; see Eq.~(\ref{VolE}) below.

As a check on our counting, consider the total color charge of the quarks:
\begin{eqnarray}
&&\vec{Q}=\sum_{i=1}^{2\eta} \vec{q}_i .
\end{eqnarray}
Squaring on both sides, we get
\begin{equation}
\begin{aligned}
\vec{Q}^2  & =\sum_{i=1}^{2\eta}{\vec{q}_i}^{\,2} +2 \sum_{i\ne j}\vec{q}_i\cdot\vec{q}_j \\ &
=2\eta\times\frac{4}{3}g^2 + 2\times\frac{1}{3^{\eta}}\sum_i E_i C_i \\&
=0.
\end{aligned}
\end{equation}
This shows our model is an overall color singlet. Note that we divided by $3^{\eta}$ in the second term to average over all the possible configurations.

The TF quark model replaces the sum over particle number in particle interaction models with an integral over the density of state particle properties. We will weight interaction strengths, taken from the classical theory, by the probabilities of various interactions in the color sector, which we assume to be flavor independent. So, we need a connection between particle number and probability. We will assume as in other TF quark models that these interaction probabilities are proportional to the number of particle interaction terms. Also, treating color-combinations on an equal footing, each color now shares one-third of the total probability. Therefore, we divide the probabilities in Table~\ref{Table4} by three as shown in Table~\ref{Table1}. Also, notice that we have generated a new way of representing of probabilities using double color indices $i$ and $j$ and barred and double barred symbols for $P$. No bar is for color-color probabilities, one bar is for color-anticolor probabilities, and two bars is for anticolor-anticolor probabilities. From Table \ref{Table1}, we can see that we correctly obtain $\sum_{i\leq j}^{3}{P_{ij}+\bar{P}_{ij}}+\bar{\bar{P}}_{i j} =1$ for the sum of all the probabilities. 

\begin{table}[ht!]
\centering
\begin{tabular}{ |c|c|c|c| }
\hline
New symbol & Old symbol  & Probability values\\
\hline
   $P_{ii}$   & $\displaystyle\frac{P_1}{3}$  & $\displaystyle\frac{(\eta -1 )} { 18(2\eta -1)}$  \\
\hline
  $P_{ij}, i< j$  & $\displaystyle\frac{P_2}{3}$ & $\displaystyle\frac{\eta -1 } { 9(2\eta -1)}$\\
 \hline
  $\bar{P}_{ij},i< j$ & $\displaystyle\frac{P_3}{3}$ & $\displaystyle\frac{2(\eta -1 )}  {9(2\eta -1)}$\\
\hline
  $\bar{P}_{ii}$ &  $\displaystyle\frac{P_4}{3}$&$\displaystyle\frac{(\eta +2 )}{  9(2\eta -1)}$\\
 \hline
  $\bar{\bar P}_{ii}$ &$\displaystyle\frac{P_5}{3}$& $\displaystyle\frac{(\eta -1 )} {18(2\eta -1)}$\\
\hline
   $\bar{\bar P}_{ij},i<j$ & $\displaystyle\frac{P_6}{3}$&$\displaystyle\frac{(\eta -1 )}  {9(2\eta -1)}$\\
 \hline
 \end{tabular}
 \caption{The coupling constants and probabilities for certain types of quark and antiquark interactions in mesons.}
\label{Table1}
\end{table}

\section{System energies and equations}\label{sec3}

Our multi quark-pair system consists of an equal number of $\eta$ quarks and $\eta$ anti-quarks. Each quark or antiquark carries both color and flavor. So, let us introduce $N^I$ as the number of quarks with flavor index $I$, and $\bar{N}^I$ as number of anti-quarks with anti-flavor index $I$ such that
\begin{equation}\label{ab1025}
\sum_{I}g^I N^I =\eta,
\end{equation}
and
\begin{equation}\label{ab102225}
    \sum_{I}\bar{g}^I\bar{N}^I =\eta.
\end{equation}
In terms of quark density these equations can be expressed as
\begin{equation}\label{ab1034}
    \sum_I \int d^3r\, n^{I}_i(r) =\eta /3,
\end{equation}
and
\begin{equation}\label{ab102226}
   \sum_I \int d^3r\, {\bar{n}}^{I}_i(r) =\eta /3.
\end{equation}
In Eqs.(\ref{ab1034}) and (\ref{ab102226}) degeneracy factors are already included in quark densities $n^{I}_i(r)$ and 
${\bar{n}}^{I}_i(r)$. 

Similarly, the sum over the color index gives the total number for a given flavor. Thus
\begin{equation}\label{ab1044}
    \sum_i \int d^3r\, n^{I}_i (r) = N^{I}g^{I}
\end{equation}
and
\begin{equation}\label{ab102227}
    \sum_i \int d^3r\, {\bar{n}}^{I}_i (r) ={\bar{N}}^{I}{\bar{g}}^{I}
\end{equation}
Also, for the convenience, we will introduce the single-particle normalized density
\begin{equation}
    {\hat{n}}^{I}_i \equiv \frac{3n^{I}_i}{N^{I}}  
    \end{equation}
    and
\begin{equation}
{\hat{\bar{n}}}^{I}_i \equiv \frac{3\bar{n}^{I}_i}{\bar{N}^{I}}.
\end{equation}
This form of quark density will be helpful in correctly normalizing the TF interaction energy when continuum sources are used.

In the earlier section, we calculated probability for six types of interactions between colors. We will now see how these color interactions are associated with flavor numbers.

For a  quark-antiquark system with $2\eta$ total number of particles, the number of interactions possible is $\eta(2\eta -1)$. Out of these $\textbf{CCA}$ and $\textbf{CCR}$ types occur \[\left( \sum_I \frac{N^{I}\left( N^{I}-1\right)}{2}(g^{I})^2 + \sum_{I\neq J}\frac{N^{I} N^{J}}{2}g^{I}g^{J} + \sum_IN^{I}\frac{g^{I}(g^{I}-1)}{2}\right) \] times,
$\textbf{AAR}$ and $\textbf{AAA}$ occurs \[\left( \sum_I \frac{\bar{N}^{I}\left( \bar{N}^{I}-1\right)}{2}(\bar{g}^{I})^2 + \sum_{I\neq J}\frac{\bar{N}^{I} \bar{N}^{J}}{2}\bar{g}^{I}\bar{g}^{J} + \sum_{I}{\bar{N}}^{I}\frac{\bar{g}^{I}(\bar{g}^{I}-1)}{2}\right) \] times,
and $\textbf{CAR}$ and $\textbf{CAA}$ occurs \[ \left( \sum_{I,J}\bar{N}^{I} N^{J} \bar{g}^{I}g^{J} \right) \] times. 
For a consistency check, we are going to add all of these seven terms. We begin with the three $\textbf{CCA}$ and $\textbf{CCR}$ terms, for which we have
\begin {equation}\label{furr}
\begin{aligned}
\quad \sum_I \frac{N^{I}\left( N^{I}-1\right)}{2}(g^{I})^2 + &\sum_{I\neq J}\frac{N^{I} N^{J}}{2}g^{I}g^{J} + \sum_IN^{I}\frac{g^{I}(g^{I}-1)}{2}\\&
= \sum_I \frac{N^{I}g^{I}}{2} \left( N^{I}g^{I} -1\right) + \sum_{I\neq J}\frac{N^{I} N^{J}}{2}g^{I}g^{J} \\&
= \sum_I \frac{\left(N^{I}g^{I}\right)^2}{2}  -\sum_I \frac{\left(N^{I}g^{I}\right)}{2} +\sum_{I\neq J}\frac{N^{I} N^{J}}{2}g^{I}g^{J} \\&
=\frac{1}{2} \left(\sum_I N^{I}g^{I}\right)^2  -\sum_I \frac{\left(N^{I}g^{I}\right)}{2} \\&
=\frac{1}{2}\left(\eta^2-\eta\right).
 \end{aligned}
\end{equation}
The interaction between antiflavors, i.e, the $\textbf{AAR}$ and $\textbf{AAA}$ terms, also yields $\eta^2/2 -\eta/2$. The last term involving one flavor and one anti-flavor gives $\eta^2$. Adding all of them yields $\eta(2\eta-1)$ as it should.

We will now use this information to write expressions for kinetic and potential energies. Building up the expression from Section~\ref{sec1}, we can write
\begin{equation}\label{kkenergy}
    \begin{aligned}
    T=&\sum_{i,I}\int^{r_{max}}{d^3r\,\frac{\left(6\pi^2\hbar^3N^{I}\hat{n}_i^{I}(r)    \right)^{5/3}}{20\pi^2\hbar^3m^{I}(g^{I})^{2/3}} +}\sum_{i,I}\int^{r_{max}}{d^3r\,\frac{\left(6\pi^2\hbar^3\bar{N}^{I}\hat{\bar{n}}_i^{I}(r)    \right)^{5/3}}{20\pi^2\hbar^3\bar{m}^{I}(\bar{g}^{I})^{2/3}} }.
    \end{aligned}
\end{equation}
for the kinetic energy from quarks and antiquarks. Here we have assumed that radius of the objects are finite. Similarly, assigning the probabilities and couplings from the interaction terms shown in Table~\ref{Table1}, we can now define total potential energy to be
\begin{equation}\label{ppenergy}
\begin{aligned}
&U= \frac{4}{3}g^2\times 
\\ &
\left\{\sum_I{\frac{N^{I}\left( N^{I}-1\right)}{2}\int{\int{d^3r\,d^3r'\frac{\left(  \sum_i{P_{ii}\hat{n}_i^{I}{(r)}\hat{n}_i^{I}{(r')}} -\frac{1}{2}\sum_{i<j}{P_{ij}\hat{n}_i^{I}{(r)}\hat{n}_j^{I}{(r')}}  \right)}{|\vec{r}-\vec{r}\,'|}}}} \right.
\\ &
+\sum_{I\neq J}{\frac{N^{I} N^{J}}{2}\int{\int{d^3r\,d^3r'\frac{\left(  \sum_i{P_{ii}\hat{n}_i^{I}{(r)}\hat{n}_i^{J}{(r')}} -\frac{1}{2}\sum_{i<j}{P_{ij}\hat{n}_i^{I}{(r)}\hat{n}_j^{J}{(r')}}  \right)}{|\vec{r}-\vec{r}\,'|}}}}
\\ &
+\sum_I{N^{I}\frac{g^{I}\left( g^{I}-1\right)}{2{\left(g^{I}\right)}^2}\int{\int{d^3r\,d^3r'\frac{\left(  \sum_i{P_{ii}\hat{n}_i^{I}{(r)}\hat{n}_i^{I}{(r')}} -\frac{1}{2}\sum_{i<j}{P_{ij}\hat{n}_i^{I}{(r)}\hat{n}_j^{I}{(r')}}  \right)}{|\vec{r}-\vec{r}\,'|}}}}
\\ &
+\sum_{{I}}{\frac{\bar{N}^{I}\left( \bar{N}^{I}-1\right)}{2}\int{\int{d^3r\,d^3r'\frac{\left(  \sum_i{\bar{\bar{P}}_{ii}\hat{\bar{n}}_i^{I}{(r)}\hat{\bar{n}}_i^{I}{(r')}} -\frac{1}{2}\sum_{i<j}{\bar{\bar{P}}_{ij}\hat{\bar{n}}_i^{I}{(r)}\hat{\bar{n}}_j^{I}{(r')}}  \right)}{|\vec{r}-\vec{r}\,'|}}}}
\\ &
+\sum_{{I}\neq {J}}{\frac{\bar{N}^{I} \bar{N}^{J}}{2}\int{\int{d^3r\,d^3r'\frac{\left(  \sum_i{\bar{\bar{P}}_{ii}\hat{\bar{n}}_i^{I}{(r)}\hat{\bar{n}}_i^{J}{(r')}} -\frac{1}{2}\sum_{i<j}{\bar{\bar{P}}_{ij}\hat{\bar{n}}_i^{I}{(r)}\hat{\bar{n}}_j^{J}{(r')}}  \right)}{|\vec{r}-\vec{r}\,'|}}}}
\\ &
+\sum_{{I}}{\bar{N}^{I}\frac{\bar{g}^{I}\left( \bar{g}^{I}-1\right)}{2({\bar{g}^{I}})^2}\int{\int{d^3r\,d^3r'\frac{\left(  \sum_i{\bar{\bar{P}}_{ii}\hat{\bar{n}}_i^{I}{(r)}\hat{\bar{n}}_i^{I}{(r')}} -\frac{1}{2}\sum_{i<j}{\bar{\bar{P}}_{ij}\hat{\bar{n}}_i^{I}{(r)}\hat{\bar{n}}_j^{I}{(r')}}  \right)}{|\vec{r}-\vec{r}\,'|}}}} 
\\ &
-\left.\sum_{{I},J}{{\bar{N}^{I} N^{J}}\int{\int{d^3r\,d^3r'\frac{\left(  \sum_i{\bar{P}_{ii}\hat{\bar{n}}_i^{I}{(r)}\hat{n}_i^{J}{(r')}} -\frac{1}{2}\sum_{i<j}{\bar{P}_{ij}\hat{\bar{n}}_i^{I}{(r)}\hat{n}_j^{J}{(r')}}  \right)}{|\vec{r}-\vec{r}\,'|}}}}\right\}.
\end{aligned}
\end{equation}
In Eq.~(\ref{ppenergy}) $\hat{n}_i^{I}$ are number densities, which are normalized to one when integrated over space\cite{WW1}. Note that the degeneracy factors are already contained in the expression for number densities. Therefore, we divided the third and sixth terms by the square of degeneracy factors.

From Table~\ref{Table1}, we can see that the probability of interaction type CCA is twice the CCR type. Also, interaction probability AAA is twice the AAR type. This simple finding amazingly removes six out of seven terms from the interaction energy, leaving us with the last term from (\ref{ppenergy}). Thus we conclude, for mesons on average, quarks only interact with antiquarks. This cancellation  makes our analytical solutions easier to achieve.

The total energy $E$ can now be written as the sum of kinetic and potential energies such that
\begin{equation}
\begin{aligned}
&E=\sum_{i,I}\int^{r_{max}}{d^3r\frac{\left(6\pi^2\hbar^3N^{I}\hat{n}_i^{I}(r)  \right)^{5/3}}{20\pi^2\hbar^3m^{I}(g^{I})^{2/3}} +}\sum_{i,I}\int^{r_{max}}{d^3r\frac{\left(6\pi^2\hbar^3\bar{N}^{I}\hat{\bar{n}}_i^{I}(r)    \right)^{5/3}}{20\pi^2\hbar^3\bar{m}^{I}(\bar{g}^{I})^{2/3}} }
 \\ &
-\frac{4}{3}g^2\sum_{{I},J}{{\bar{N}^{I} N^{J}}\int{\int{d^3r\,d^3r'\frac{\left(  \sum_i{\bar{P}_{ii}\hat{\bar{n}}_i^{I}{(r)}\hat{n}_i^{J}{(r')}} -\frac{1}{2}\sum_{i<j}{\bar{P}_{ij}\hat{\bar{n}}_i^{I}{(r)}\hat{n}_j^{J}{(r')}}  \right)}{|\vec{r}-\vec{r}\,'|}}}}.
\end{aligned}
\end{equation}
Now, we switch back to normalization $n^{I}_{i}$ and $\bar{n}^{{I}}_{i}$ and assume equal Fermi color momenta, $n^{I}\equiv n^{I}_{1} \equiv n^{I}_{2} \equiv n^{I}_{3}$ for each I. The same assumption is made for antiparticles. This gives
\begin{equation}\label{energiee}
 \begin{aligned}
&E=\sum_{I} \int^{r_{max}}
{
d^3r \frac {3\left( 6\pi^2\hbar^3N^{I}{n}^{I}(r) \right)^{5/3} }{20\pi^2\hbar^3m^{I}(g^{I})^{2/3}}
+   }\\&
\sum_{I}\int^{r_{max}}
{
d^3r\frac {3\left(6\pi^2\hbar^3\bar{N}^{{I}}{\bar{n}}^{{I}}(r)    \right)^{5/3}} { 20\pi^2\hbar^3\bar{m}^{{I}}(\bar{g}^{I})^{2/3} }
}
- \frac{9\times4/3 g^2}{2\eta -1} \sum_{{I},J}\int \int \frac{d^3r d^3r'}{|\vec{r}-\vec{r}\,'|}
\bar{n}^{{I}} (r) n^{J}(r') .
 \end{aligned}
 \end{equation}
Eqs.~(\ref{ab1044}) and (\ref{ab102227}) can now be averaged over colors as 
\begin{equation}\label{ab1061}
   \int d^3r\, n^{I}(r) = N^{I}g^{I}/3,
\end{equation}
and
\begin{equation}\label{abc1061}
    \int d^3r\,{\bar{n}}^{I}(r) = {\bar{N}}^{I}{\bar{g}}^{I}/3.
\end{equation}
We will use this pair of equations to set up the normalization conditions.

\section{Thomas Fermi Quark Equations}\label{sec4}

We can now introduce Lagrange's undetermined multipliers $\lambda^{I}$ and $\bar{\lambda}^{{I}}$ associated with constraints in Eqs.(\ref{ab1061}) and (\ref{abc1061}) and add them to the expression for energy. This gives
\begin{equation}
 \begin{aligned}
  E=&\sum_{I} \int^{r_{max}}
  {
  d^3r \frac {3\left( 6\pi^2\hbar^3N^{I}{n}^{I}(r) \right)^{5/3} }{20\pi^2\hbar^3m^{I}(g^{I})^{2/3}}
  +   }
  \sum_{I}\int^{r_{max}}
  {
  d^3r\frac {3\left(6\pi^2\hbar^3\bar{N}^{{I}}{\bar{n}}^{{I}}(r)    \right)^{5/3}} { 20\pi^2\hbar^3\bar{m}^{{I}} (\bar{g}^{I})^{2/3}}
  }
  \\ &
  - \frac{9\times4/3 g^2}{2\eta -1} \sum_{{I},J}\int \int \frac{d^3r d^3r'}{|\vec{r}-\vec{r}\,'|}
  \bar{n}^{{I}}    (r) n^{J}(r') 
 \\ &
 + \sum_I \lambda^{I}\left( 3\int d^3r \,n^{I}(r)-N^{I}g^{I} \right)
 + \sum_{{I}} \bar{\lambda}^{{I}}\left( 3\int d^3r\,\bar{n}^{{I}}(r) -\bar{N}^{{I}}\bar{g}^{{I}} \right).
 \end{aligned}
 \end{equation}
The purpose of adding these terms involving Lagrange multipliers is to allow a minimization of the total energy while keeping particle number constant. The density variations $\delta \bar{n}^I (r)$ and $\delta n^{{I}}(r)$ give
\begin{equation}\label{ab622}
 \frac{(6 { \pi^2} {\hbar^3})^{5/3}}  {\pi^2 \hbar^3}   \left[\frac{1}{4\bar{m}^{{I}}}\left(\frac{\bar{n}^{{I}}(r)}{g^{{I}}}\right)^{2/3}\right] = -3 \bar{\lambda}^{{I}} + \frac{9 \times \frac{4}{3} g^2}{(2 \eta -1)} \sum_{ I} {\int \frac{ d^3 r'}{|\vec{r}-\vec{r}\,'|} {{n}}^{{I}} {(r')} }.
 \end{equation}
 \begin{equation}\label{ab1020}
 \frac{(6 { \pi^2} {\hbar^3})^{5/3}}  {\pi^2 \hbar^3}  \left [\frac{1}{4m^I}\left(\frac{n^I(r)}{g^{I}}\right)^{2/3}\right] = -3 \lambda^I + \frac{9 \times \frac{4}{3} g^2}{(2 \eta -1)} \sum_{I} {\int \frac{ d^3 r'}{|\vec{r}-\vec{r}\,'|} {\bar{n}}^{{I}} {(r')} },
 \end{equation}
 
Assuming spherical symmetry, the TF spatial functions $f^{I}(r)$ and $\bar{f}^{I}(r)$ are defined such that
\begin{equation}\label{fn3}
f^{I}(r)\equiv \frac{ra}{({ 8\alpha_{s}}/{3})} \left( \frac{6\pi^{2}n^{I}(r)}{g^{I}} \right)^{2/3},
\end{equation}
\begin{equation}\label{fn1}
\bar{f}^{I}(r)\equiv \frac{ra}{(8\alpha_{s}/3)}\left( \frac {6\pi^{2}\bar n^{I}(r)}{\bar{g}^{I}} \right)^{2/3},
\end{equation}
where
\begin{equation}\label{mattt}
a\equiv \frac{\hbar}{ m^1 c},
\end{equation}
gives the scale, where $m^1$ is the mass of lightest quark, and
\begin{equation}\label{ratt}
\alpha_{s}=\frac{g^2}{\hbar c},
\end{equation}
is the strong coupling constant. Equation (\ref{ab622}) after using Eqs.(\ref{fn3}) - (\ref{ratt}) can now be written as
\begin{equation}\label{ab1063}
 \begin{aligned}
 \frac{3m^1}{\bar{m}^I} \frac{4}{3}g^2 \frac{{\bar{f}}^{{I}}(r)}{r} = & - 3 \bar{\lambda}^{{I}} + \frac{6\cdot\frac{4}{3}g^2}{(2\eta-1)\pi} \left(\frac{2\times\frac{4}{3}\alpha_s}{a}\right)^{3/2} \\ & \times\sum_I g^{I} \left[  \frac{1}{r}\int_0^{r} dr'{r'}^2\left(\frac{f^I(r')}{r'}\right)^{3/2}   +\int_0^{r_{m}}dr'r'\left(\frac{f^I(r')}{r'}\right)^{3/2} \right].
  \end{aligned}
 \end{equation}
Here we have used the integral
\begin{equation}
    \int^{r_{max}} d^3r'\frac{n^{I}(r')}{|\vec{r}-\vec{r}\,'|} = 4\pi \left[ \int_0^r dr' {r'}^2 \frac{n^{I}(r')}{r}+\int_r^{rmax} dr' {r'}^2 \frac{n^{I}(r')}{r'}\right],
  \end{equation}
resulting from spherical symmetry. Equation (\ref{ab1063}) can be further simplified into
\begin{equation}\label{abb633}
\begin{aligned}
\bar{\alpha}^I {\bar{f}}^{{I}} (r) =\frac{-{\bar{\lambda}}^{{I}} r}{\frac{4}{3}g^2} +&\frac{2}{(2\eta-1)\pi}\left(\frac{2\times\frac{4}{3}\alpha_s}{a}\right)^{3/2}\\ & \times\sum_I g^{I}\left[\int_0^{r} dr'r'^2\left(\frac{f^I(r')}{r'}\right)^{3/2}  +r\int_0^{r_{m}}dr'r'\left(\frac{f^I(r')}{r'}\right)^{3/2}\right].
\end{aligned}
\end{equation}
At this stage we introduce the mass ratios
\begin{equation}\label{ratio}
\alpha^I \equiv \frac{m^1}{m^I}, \,\,\bar{\alpha}^I \equiv \frac{m^1}{\bar{m}^I},
\end{equation}
for both the particle and antiparticle cases.

Let us introduce the dimensionless parameter $x$ such that $r=Rx$ where
\begin{equation}\label{fn2}
R\equiv \frac{a}{(8\alpha_{s}/3)}\left(\frac{3\pi\eta }{2}\right)^{2/3}.
\end{equation}
Eq.~(\ref{abb633}) can now be written as
\begin{equation}\label{ab1093}
\begin{aligned}
\bar{\alpha}^{{I}}\bar{f}^{{I}} (x)=&\frac{-\Bar{\lambda}^{{I}}Rx}{\frac{4}{3}g^2}+\\&
\frac{3\eta}{2\eta -1}\sum_I g^{I}\left[\int_0^{x} dx'\,x'^2\left(\frac{f^I(x')}{x'}\right)^{3/2}  +x\int_x^{x_{m}}dx'\,x'\left(\frac{f^I(x')}{x'}\right)^{3/2}\right].
\end{aligned}
 \end{equation}
Similarly, starting from  Eq.~(\ref{ab1020}) we obtain the other TF integral equation:
\begin{equation}\label{ab1100}
\begin{aligned}
 \alpha^I f^I {(x)}=&\frac{-\lambda^{I}Rx}{\frac{4}{3}g^2}+\\&
 \frac{3\eta}{2\eta -1}\sum_{{I}} \bar{g}^{{I}}\left[\int_0^{x} dx'\,{x'}^2 \left(\frac{\bar{f}^{{I}}(x')}{x'}\right)^{3/2}+ x\int_x^{x_{m}} dx'\,x'\left(\frac{\bar{f}^{{I}}(x')}{x'}\right)^{3/2}\right].
 \end{aligned}
 \end{equation}
Taking first derivatives of Eqs.~(\ref{ab1093}) and (\ref{ab1100}), we have
\begin{equation}
\bar{\alpha}^I \frac{d{{\bar{f}}^I}(x)}{dx} = \frac{-{\bar{\lambda}}^{{I}}{R}}{\frac{4}{3}g^2} +\frac{3\eta}{(2\eta-1)}\left[\sum_{{I}}g^{I}\int_x^{x_{max}}dx'\,x'\left(\frac{{{f}^{I}}(x')}{x'}\right)^{3/2}\right],
 \end{equation}
and
\begin{equation}
 \alpha^I \frac{d{f^I}(x)}{dx} = \frac{-\lambda^I{R}}{\frac{4}{3}g^2} +\frac{3\eta}{(2\eta-1)}\left [\sum_{{I}}\bar{g}^{{I}}\int_x^{x_{m}}dx'\,x' \left(\frac{\bar{f}^{{I}}(x')}{x'}\right)^{3/2}\right].
 \end{equation}
Similarly, second derivatives of Eqs.~(\ref{ab1093}) and (\ref{ab1100}) yield
\begin{equation}\label{TF2}
\bar{\alpha}^{ I} \frac {d^2\bar{f}^{ I}(x)}{dx^2} = - \frac {3\eta} {(2\eta -1)}\frac{1}{\sqrt{x}} \sum_{{I}} g^{I}{ {f}^{{I}}(x)}^{3/2},
\end{equation}
and
\begin{equation}\label{TF1}
\alpha ^I \frac {d^2f^{I}(x)}{dx^2} = - \frac {3\eta} {(2\eta -1)}\frac{1}{\sqrt{x}} \sum_{{I}} \bar{g}^{{I}}{ \bar{f}^{{I}}(x)}^{3/2}.
\end{equation}

Equations (\ref{TF2}) and (\ref{TF1}) are the differential form of the TF quark equations in the case of mesons. The interchangeability of these equations shows the TF equations are invariant with respect to particle and antiparticle. As was mentioned before, it also shows that quarks interact only with antiquarks in mesonic matter; quark/quark and antiquark/antiquark interactions sum to zero in the TF model. When there is an explicit particle/antiparticle symmetry, we assume $f=\bar f$ to reduce the TF differential equations into one incredibly simple form:
\begin{equation}\label{tf1}
\alpha ^I \frac {d^2f^{I}(x)}{dx^2} = - \frac {3\eta} {(2\eta -1)} \frac{1}{\sqrt{x}}\sum_{I} g^{I}{ f^{I}(x)}^{3/2}.
\end{equation}
We use this equation to form system energies and equations for charmonium/bottomonium families and $Z$-meson family systems. However, for $D$ and $B$-meson families, the explicit asymmetry in the masses of the particle and antiparticle will require us take a different approach.

\section{Energy equations in terms of dimensionless radius}\label{sec5}

The expressions for kinetic and potential energies now need to be given in terms of the dimensionless distance in order for us to be able to obtain analytical solutions. In addition, the volume energy term that produces an external pressure on the system needs to be appropriately introduced.

To obtain the kinetic and potential energy expressions, we begin from Eq.~(\ref{energiee}) and apply Eqs.~(\ref{fn3}), (\ref{fn1}) and (\ref{fn2}). This gives the kinetic energy as
\begin{equation}\label{Keke}
\begin{aligned}
T=& \sum_{I}{\frac{12}{5\pi}\left( \frac {3\pi \eta}{2}\right)^{1/3} \frac {\frac{4}{3}g^2 \cdot \frac{4}{3}\alpha_s}{a} }  {\alpha_I g^{I} \int_{0}^{x_I}dx{\frac{\left({f}^{{I}}(x)\right)^{5/2}}{\sqrt {x}}}} \\ &
+ \sum_{I}{\frac{12}{5\pi}\left( \frac {3\pi \eta}{2}\right)^{1/3} \frac {\frac{4}{3}g^2 \cdot \frac{4}{3}\alpha_s}{a} }  {\bar{\alpha}_I \bar{g}^{I} \int_{0}^{x_I}dx{\frac{\left({\bar{f}}^{{I}}(x)\right)^{5/2}}{\sqrt {x}}}}.
\end{aligned}
\end{equation}
The potential energy
\begin{equation}
\begin{aligned}
U =& - \frac{9\cdot \frac{4}{3}g^2}{\left( 2\eta -1 \right)}   \sum_{I,J}\bar{g}^{I}g^{J}{\int^{r_I}\int^{r_J} d^3r \, d^3r'  \frac {n^{I}(r) n^{J}(r')}{|\vec{r}-\vec{r}\,'|}},
\end{aligned}
\end {equation}
becomes
\begin{equation}\label{Ufinal}
\begin{aligned}
&U=\\
&-\frac{9\cdot\frac{4}{3}g^2}{\left( 2\eta -1 \right)}\frac{\eta^{2}}{R}  \sum_{I,J}{\quad \bar{g}^{I}g^{J}\Big[   \int_0^{x_I}{dx \frac{\left({\bar{f}}^{{I}}(x)\right)^{3/2}}{\sqrt{x}} \int_0^{x}{dx' \sqrt{x'}\left({f}^{{J}}(x')\right)^{3/2}}}}\\ & +  \int_0^{x_I}{dx {\left({\bar{f}}^{{I}}(x)\right)^{3/2}}{\sqrt{x}} \int_x^{x_J}{dx'  \frac{\left({f}^{{J}}(x')\right)^{3/2}}{\sqrt{x'}}}} \Big].
\end{aligned}
\end{equation}

The volume energy ($E_v$) term gives the inward pressure from the vacuum. We assume that\cite{Mit} 
\begin{equation}\label{VolE}
\begin{aligned}
E_v= \frac{4}{3}\pi R^3 x_{max}^3  B,
\end{aligned}
\end{equation}
where $B$ is the bag constant. Now that we have $T$, $U$ and $E_v$  terms, we can find the total energy of a desired multi quark-pair state. The total energy of such a state is simply given by
\begin{equation}\label{Emass}
\begin{aligned}
E= T+ V+ E_v + \eta \cdot m_q +\eta \cdot \bar{m}_q,
\end{aligned}
\end{equation}
where $m_q$ and $\bar{m}_q$ are the mass of the quark and anti-quark respectively.

In Section~\ref{sec7} we will fit model parameters for a given set of mesons, including $m_q$ and $\bar{m}_q$. However, to assess the stability of multi quark-pair mesons we omit the trivial mass part and will examine the energy as a function of quark content.

\section{Application of model}\label{sec6}
In this section we will now apply our model and obtain energy expressions for different families of quarks. In all the cases that follow charm can be read as bottom also.


\subsection{Charmonium family: Case 1}
This type of family contains quarks and anti-quarks of equal mass. The system of quarks in this family can be represented as $Q\bar{Q}, Q\bar{Q}Q\bar{Q}, Q\bar{Q}Q\bar{Q}Q\bar{Q} $ and so on. We have investigated the case where $Q$ is a charm quark or bottom quark but not both in the same system. We call it the charmonium family. Charmonium and all the multi quark-pair families of charmonium consists of charm and anti-charm (or bottom and anti-bottom) only. Therefore, $I$ takes a single value and hence is dropped. $g_0$ refers to degeneracy and, for this application, can have the value of one or two to represent spin. This can give spin splittings, but we have not yet included explicit spin-spin interactions in the model, like we have done in our baryon model\cite{WW2}. Eq.~(\ref{tf1}) becomes
\begin{equation}
\begin{aligned}
\frac {d^2f(x)}{dx^2} = - \frac {3\eta} {(2\eta -1)} \cdot g_0\cdot \frac{1}{\sqrt{x}}{ {f}{(x)}}^{3/2},
\end{aligned}
\end{equation}
where
\begin{equation}
g_0 \times N^{I}=\eta.
\end{equation}
We choose the normalization equation to be
\begin{equation}
\int_0^{x_{max}}dx{\sqrt{x}} { {f}{(x)}}^{3/2} =\frac {N^{I}}{3\eta}.
\end{equation}
Expressing the normalization equation in terms of boundary conditions, we have
\begin{equation}
\bigl(x\frac{df}{dx}-f \bigr)|_{x_{max}}=-\frac{\eta}{2\eta -1}.
\end{equation}

With these modified boundary conditions, we can derive the expressions for kinetic and potential energies. For the kinetic energy we can start with Eq.~(\ref{Keke}) and assume a single flavor. In this case, both quarks and antiquarks contribute equally to the kinetic energy, which gives
\begin{equation}\label{Kekey}
\begin{aligned}
T= 2\cdot{\frac{12}{5\pi}\left( \frac {3\pi \eta}{2}\right)^{1/3} \frac {\frac{4}{3}g^2 \cdot \frac{4}{3}\alpha_s}{a} }  { g_0 \int_{0}^{x_{max}}dx{\frac{\left(f(x)\right)^{5/2}}{\sqrt {x}}}}.
\end{aligned}
\end{equation}
The integral may be done using the TF differential equation, and results in
\begin{equation}
\begin{aligned}
&T= \frac{24}{5\pi}\left( \frac {3\pi \eta}{2}\right)^{1/3} \frac {\frac{4}{3}g^2 \cdot \frac{4}{3}\alpha_s}{a}  \Bigg[ -\frac{5}{21} \frac {df(x)}{dx}|_{x_{max}} + \frac {4}{7} \sqrt{x_{max}}\left({f}(x_{max})\right)^{5/2} g_0   \Bigg].
\end{aligned}
\end {equation}
Thus, the kinetic energy depends only on the derivative and value of the TF function at the boundary. When there is a single flavor as in the case of the charmonium family, and symmetry of particle and antiparticle occurs, the potential energy  Eq.~(\ref{Ufinal}) can be written as
\begin{equation}\label{Ufinale}
\begin{aligned}
U=-\frac{9\cdot\frac{4}{3}g^2}{\left( 2\eta -1 \right)}\frac{\eta^{2}}{R} & { \left({g}^{I}\right)^2 \Big[   \int_0^{x_{max}}{dx \frac{\left({{f}}(x)\right)^{3/2}}{\sqrt{x}} \int_0^{x}{dx' \sqrt{x'}\left({f}(x')\right)^{3/2}}}}\\ & +  \int_0^{x_{max}}{dx {\left({{f}}(x)\right)^{3/2}}{\sqrt{x}} \int_x^{x_{max}}{dx'  \frac{\left({f}(x')\right)^{3/2}}{\sqrt{x'}}}} \Big].
\end{aligned}
\end{equation}
This can be further simplified to
 \begin{eqnarray}
U= \frac{4}{\pi}\left( \frac {3\pi \eta}{2}\right)^{1/3} \frac {\frac{4}{3}g^2 \cdot \frac{4}{3}\alpha_s}{a}\times   \Bigg[ \frac{4}{7} \frac {df(x)}{dx}|_{x_{max}} - \frac {4}{7} \sqrt{x_{max}}\left({f}(x_{max})\right)^{5/2} g_0   \Bigg],
\end {eqnarray}
which like the expression for $T$ depends on the derivative and values of the TF function at the boundary. It is important to realize that the \lq\lq a" value in this section differs from others where the lightest quark is not charm or bottom.


\subsection{$Z$-meson family: Case 2}
The constituents of $Z$-mesons are charm ($c$) (or bottom($b$)), anti-charm ($\bar{c}$) (or anti-bottom ($\bar{b}$)), light ($u$ or $d$) and anti-light quarks. The particles in this family can be represented as $Q\bar{Q}q\bar{q}$, $Q\bar{Q}q\bar{q}Q\bar{Q}q\bar{q}$, $Q\bar{Q}q\bar{q}Q\bar{Q}q\bar{q}Q\bar{Q}q\bar{q}$, and so on, where $Q$ represents a heavy quark and $q$ is a light quark. We treat the mass of the up and down quarks as the same. This means the $Z$-meson and all multi quark-pair families of $Z$-mesons have a total quark mass equal to the antiquark mass. As before, we set $f=\bar{f}$ in the TF equations and obtain Eq.~(\ref{tf1}) with $I=1$ or 2. Let $f^{1}(x)$ be the TF function of the light quark and $f^{2}(x)$ be the TF function of the heavier quark. For $N_{1}$ quarks with degeneracy factor $g_1$, and $N_2$ quarks with degeneracy $g_2$, we have
\begin{eqnarray}
&& g_1 N_1 + g_2 N_2 =\eta.
 \end{eqnarray}
In our application $g_1$ can have values 1, 2, 3, or 4 whereas $g_2$ can have value of 1 or 2 only. We assume a linear relation exists between $f^{1}(x)$ and $f^{2}(x)$ in the region $0<x<x_{2}$, and that $f^{2}(x)$ vanishes for a dimensionless distance greater than $x_{2}$, i.e.,
\begin{eqnarray}
\begin{aligned}
&f^{1}(x) = k f^{2}(x)  \quad \text{for} \quad 0\leq x \leq x_2,  \\
& f^{2}(x) = 0       \qquad \text{for} \quad x_2 \leq x \leq x_1.
\end{aligned}
\end{eqnarray}
Eq.~(\ref{tf1}) can now be written as two set of equations:
\begin{equation}\label{zz2}
\begin{aligned}
\alpha ^1 \frac {d^2f^{1}(x)}{dx^2} = - \frac {3\eta} {(2\eta -1)} \frac{1}{\sqrt{x}} \bigl( g_1\left({f}^{{1}}(x)\right)^{3/2}  + g_2\left({f}^{{2}}(x)\right)^{3/2}  \bigr),
\end{aligned}
\end{equation}
\begin{equation}\label{zz3}
\begin{aligned}
\alpha ^2 \frac {d^2f^{2}(x)}{dx^2} =- \frac {3\eta} {(2\eta -1)}\frac{1}{\sqrt{x}}  \Big( g_1\left({f}^{{1}}(x)\right)^{3/2}  +g_2\left({f}^{{2}}(x)\right)^{3/2}  \Big).
\end{aligned}
\end{equation}
To make Eqs.~(\ref{zz2}) and (\ref{zz3}) consistent, we need
\begin{equation}
k = \frac{\alpha ^2}{\alpha ^1},
\end{equation}
which is just the inverse ratio of the given masses from (\ref{ratio}). The similar step in the case of baryons gives a much more complicated consistency condition\cite{WW2}. The normalization conditions are
\begin{equation}
\int_{0}^{x_2}{x^{1/2}\left({f}^{{2}}(x)\right)^{3/2} dx} = \frac {N_2}{3\eta},
\end{equation}
and
\begin{equation}
\int_{0}^{x_1}{x^{1/2}\left({f}^{{1}}(x)\right)^{3/2} dx} = \frac {N_1}{3\eta}.
\end{equation}
In region $0<x<x_{2}$,
\begin{equation}
 \frac {d^2f^{1}(x)}{dx^2} = Q_1 \frac {\left({f}^{{1}}(x)\right)^{3/2} }{\sqrt{x}},
\end {equation}
where
\begin{equation}
Q_1 = - \frac{3\eta}{\left( 2\eta -1\right)\alpha_1}\left( g_1 + \frac {g_2}{k^{ 3/2}}  \right).
\end {equation}
In region $x_2<x<x_1$,
\begin{equation}
\frac {d^2f^{1}(x)}{dx^2} = Q_2 \frac {\left({f}^{{1}}(x)\right)^{3/2} }{\sqrt{x}},
\end {equation}
where
\begin{equation}
Q_2 = - \frac{3\eta}{\left( 2\eta -1\right)\alpha_1}g_1 .
\end {equation}
With the equations above we can express normalization conditions in the form of boundary conditions:
\begin{equation}
\left.\left(   x\frac {df^{2}(x)}{dx} - f^{2}(x)   \right)\right|_{x_{2}} = - \frac {N_2 Q_1 k^{3/2}}{3\eta},
\end {equation}
\begin{equation}
\left.\left(   x\frac {df^{1}(x)}{dx} - f^{1}(x)   \right)\right|_{x_{1}} = - \frac {\eta}{\left( 2\eta -1 \right) \alpha_{1}}.
\end {equation}

The energies are derived as for Case 1. For $g_1$ flavors with $N_1$ particles and $g_2$ flavors with $N_2$ particles we have
\begin{eqnarray}
\begin{aligned}
T= 2\cdot\frac{12}{5\pi}\left( \frac {3\pi \eta}{2}\right)^{1/3} \frac {\frac{4}{3}g^2 \cdot \frac{4}{3}\alpha_s}{a}  &\Bigg[ g_1 \alpha_1 \int_{0}^{x_1}{\frac{\left({f}^{{1}}(x)\right)^{5/2}}{\sqrt {x}}dx}+\\
& g_2 \alpha_2 \int_{0}^{x_2}{\frac{\left({f}^{{2}}(x)\right)^{5/2}}{\sqrt {x}}dx}\Bigg].
\end{aligned}
\end {eqnarray}
Using the TF function differential equations, the consistency condition for $k$ and boundary conditions allows one to relate the integrals to TF function values and derivatives on the surfaces:
\begin{eqnarray}
\begin{aligned}
T= 2\cdot\frac{12}{5\pi}&\left( \frac {3\pi \eta}{2}\right)^{1/3} \frac {\frac{4}{3}g^2 \cdot \frac{4}{3}\alpha_s}{a}  \Bigg[ -\frac{5}{21}\alpha_1 \frac {df^1(x)}{dx}|_{x_1} \\&+ \frac {4}{7} \sqrt{x_1}\left({f}^{{1}}(x_1)\right)^{5/2} g_1 \alpha_1  +  \frac {4}{7} \sqrt{x_2}\left({f}^{{2}}(x_2)\right)^{5/2} g_2 \alpha_2  \Bigg].
\end{aligned}
\end {eqnarray}
The expression for potential energy in the same case becomes
\begin{equation}\label{Ucase2}
\begin{aligned}
U=-\frac{9\cdot\frac{4}{3}g^2}{\left( 2\eta -1 \right)}\frac{\eta^{2}}{R}& \sum_{I,J}{ {g}^{I}g^{J}  \Big[   \int_0^{x_I}{dx \frac{\left({{f}}^{{I}}(x)\right)^{3/2}}{\sqrt{x}} \int_0^{x}{dx' \sqrt{x'}\left({f}^{{J}}(x')\right)^{3/2}}}}\\ & +  \int_0^{x_I}{dx {\left({{f}}^{{I}}(x)\right)^{3/2}}{\sqrt{x}} \int_x^{x_J}{dx'  \frac{\left({f}^{{J}}(x')\right)^{3/2}}{\sqrt{x'}}}} \Big].
\end{aligned}
\end{equation}
which can be placed into the form
\begin{equation}
U=- \frac{9\cdot\frac{4}{3}g^2}{\left( 2\eta -1 \right)}\frac{\eta^{2}}{R} \Bigg[   g_1^2 K_1 + g_2^2K_2+ g_1g_2K_{12}+ g_1g_2K_{21}\Bigg] ,
\end{equation}
where
\begin{equation}
\begin{aligned}
K_1 \equiv    \int_0^{x_1}{dx \frac{\left({f}^{{1}}(x)\right)^{3/2}}{\sqrt{x}} \int_0^{x}{dx' \sqrt{x'}\left({f}^{{1}}(x')\right)^{3/2}}} +  \\
\int_0^{x_1}{dx {\left({f}^{{1}}(x)\right)^{3/2}}{\sqrt{x}} \int_x^{x_1}{dx'  \frac{\left({f}^{{1}}(x')\right)^{3/2}}{\sqrt{x'}}}},
\end{aligned}
\end{equation}
\begin{equation}
\begin{aligned}
K_2 \equiv    \int_0^{x_2}{dx \frac{\left({f}^{{2}}(x)\right)^{3/2}}{\sqrt{x}} \int_0^{x}{dx' \sqrt{x'}\left({f}^{{2}}(x')\right)^{3/2}}} + \\
 \int_0^{x_2}{dx {\left({f}^{{2}}(x)\right)^{3/2}}{\sqrt{x}} \int_x^{x_2}{dx'  \frac{\left({f}^{{2}}(x')\right)^{3/2}}{\sqrt{x'}}}} ,
\end{aligned}
\end{equation}
\begin{equation}
\begin{aligned}
K_{12}\equiv     \int_0^{x_1}{dx \frac{\left({f}^{{1}}(x)\right)^{3/2}}{\sqrt{x}} \int_0^{x}{dx' \sqrt{x'}\left({f}^{{2}}(x')\right)^{3/2}}}  +  \\
\int_0^{x_1}{dx {\left({f}^{{1}}(x)\right)^{3/2}}{\sqrt{x}} \int_x^{x_2}{dx'  \frac{\left({f}^{{2}}(x')\right)^{3/2}}{\sqrt{x'}}}} ,
\end{aligned}
\end{equation}
and
\begin{equation}
\begin{aligned}
K_{21}\equiv    \int_0^{x_2}{dx \frac{\left({f}^{{2}}(x)\right)^{3/2}}{\sqrt{x}} \int_0^{x}{dx' \sqrt{x'}\left({f}^{{1}}(x')\right)^{3/2}}}  +  \\
\int_0^{x_2}{dx {\left({f}^{{2}}(x)\right)^{3/2}}{\sqrt{x}} \int_x^{x_1}{dx'  \frac{\left({f}^{{1}}(x')\right)^{3/2}}{\sqrt{x'}}}}.
\end{aligned}
\end{equation}
One can show that the $K_{12}$ integral is equivalent to the $K_{21}$ integral. 

At this point we would like to take some time to explain the reductions of these expressions. We will see some interesting cancellations in the analytical solutions. The expressions for $K_1$, $K_2$ and $K_{12}$ after integrations and reductions can be written as
\begin{equation}
     \begin{aligned}
    K_1&= \int_0^{x_2} dx \frac{\left( f^1(x) \right)^{5/2}}{\sqrt{x}} \left(-\frac{1}{Q_1} \right) \\&
    + \int_0^{x_2} dx \sqrt{x} \left( f^1 (x) \right)^{3/2} \left( \frac{1}{Q_1}\left. \frac{df^1(x)}{dx}\right|_{x_2} + \left. \frac{1}{Q_2} \frac{df^1(x)}{dx}\right|_{x_1} -\left. \frac{1}{Q_2} \frac{df^1(x)}{dx}\right|_{x_2}  \right) \\&
    +\int_{x_2}^{x_1} dx \frac{ \left( f^1(x)\right)^{3/2} }{\sqrt{x}}  \left(   \frac{N_2 k^{3/2}}{\eta} \left( 1- \frac{Q_1}{Q_2} \right)  \right)\\&
        +\int_{x_2}^{x_1} dx \frac{ \left( f^1(x)\right)^{5/2} }{\sqrt{x}}  \left(  -\frac{1}{Q_2}  \right)\\&
            +\int_{x_2}^{x_1} dx { \left( f^1(x)\right)^{3/2} }{\sqrt{x}}  \left(  \left.  \frac{1}{Q_2} \frac{df^1(x)}{dx}\right|_{x_1}   \right),
     \end{aligned}
 \end{equation}
\begin{equation}
     \begin{aligned}
          K_2 & = \int_0^{x_2} dx \frac{\left( f^2(x) \right)^{5/2}}{\sqrt{x}}  \left(  - \frac{1}{Q_1 k^{1/2}} \right) \\&
     +\int_{0}^{x_2} dx { \left( f^2(x)\right)^{3/2} }{\sqrt{x}}   \left(  \left. \frac{1}{Q_1 k^{1/2}}  \frac{df^2(x)}{dx} \right|_{x_2} \right),
          \end{aligned}
 \end{equation}
and
  \begin{equation}
     \begin{aligned}
      K_{12}& =  \int_0^{x_2} dx \frac{\left( f^2(x) \right)^{5/2}}{\sqrt{x}}  \left(- \frac{k}{Q_1} \right) \\&
     + \int_{0}^{x_2} dx { \left( f^2(x)\right)^{5/2} }{\sqrt{x}} \left( \left.\frac{k}{Q_1}  \frac{df^2(x)}{dx} \right|_{x_2} \right)\\&
     +\int_{x_2}^{x_1} \frac{\left( f^1(x) \right)^{3/2}}{\sqrt{x}} \left( \frac{N_2}{\eta} \right).
     \end{aligned}
 \end{equation}
Similarly, the potential energy now simplifies to
 \begin{equation}
     \begin{aligned}
U&=-\frac{9\cdot\frac{4}{3}g^2}{\left( 2\eta -1 \right)}\frac{\eta^{2}}{R} \times  \\& \left [ \int_0^{x_2}dx\frac{\left(f^2(x)\right)^{5/2}}{\sqrt{x}}  \left[\frac{-k^{5/2}}{Q_1} {g_1}^2 - \frac{1}{Q_1 k^{1/2}} {g_2}^2  - \frac{k}{Q_1} 2 g_1 g_2\right] \right. \\&
      +     \int_0^{x_2}dx {\left(f^2(x)\right)^{3/2}}{\sqrt{x}} \,\, \times \\ & \left[ {g_1}^2\left(\left. \frac{k^{5/2}}{Q_1} \frac{df^2(x)}{dx}\right|_{x_2} +\left.\frac{k^{3/2}}{Q_2} \frac{df^1(x)}{dx}\right|_{x_1} - \left.\frac{k^{5/2}}{Q_2} \frac{df^2(x)}{dx}\right|_{x_2}\right) \right. \\& \left.+ \left.{g_2}^2 \frac{1}{Q_1 k^{1/2}} \frac{df^2(x)}{dx}\right|_{x_2}    +  \left.2g_1 g_2 \frac{k}{Q_1} \frac{df^2(x)}{dx}\right|_{x_2}   \right]  \\&
      +\int_{x_2}^{x_1} dx \frac{(f^1(x))^{3/2}}{\sqrt{x}} \left[ \frac{N_2k^{3/2}}{\eta}  \left( 1- \frac{Q_1}{Q_2}\right) {g_1}^2 + \frac{N_2}{\eta}2{g_1}{g_2}\right] \\&
    \left.  +\int_{x_2}^{x_1} dx {(f^1(x))^{3/2}}{\sqrt{x}} \left[\left.\frac{1}{Q_2} \frac{df^1(x)}{dx}\right|_{x_1} {g_1}^2\right] + \right.\\&
   \left.  \int_{x_2}^{x_1} \frac{\left(f^1(x)\right)^{5/2} }{\sqrt{x}} \left[-\frac{1}{Q_2} {g_1}^2\right] \right],
     \end{aligned}
     \end{equation}
which on further reduction yields
  \begin{equation}
 \begin{aligned}
U&= -\frac{9\cdot\frac{4}{3}g^2}{\left( 2\eta -1 \right)}\frac{\eta^{2}}{R} \times  \\ & \left[\left.\frac{N_2}{\eta} \frac{df^2(x)}{dx}\right|_{x_2}  \times \left[ \frac{5}{7} \frac{k^{5/2}{g_1}^2}{Q_1} + \frac{5}{7} \frac{{g_2}^2}{Q_1 k^{1/2}} + \frac{5}{7} \frac{k}{Q_1} 2{g_1g_2}  +  \frac{k^{5/2}{g_1}^2}{Q_1} -  \frac{k^{5/2}{g_1}^2}{Q_2}   \right.\right.\\& \left.  +\frac{{g_2}^2}{Q_1 k^{1/2}} + \frac{k}{Q_1} 2g_1g_2  -\frac{k^{5/2}}{Q_2} \left( 1-\frac{Q_1}{Q_2}\right) {g_1}^2 - \frac{k}{Q_2} 2{g_1g_2}  - \frac{5}{7Q_2} Q_1 k^{5/2} \frac{{g_1}^2}{Q_2}\right]  \\&
     + \frac{df^1(x)}{dx}  \times \left[ \frac{N_2}{\eta}  \frac{k^{3/2}{g_1}^2}{Q_2}+ \frac{N_2 k^{3/2}}{\eta Q_2} \left(1- \frac{Q_1}{Q_2}\right) {g_1}^2 +\frac{N_2}{\eta Q_2} 2{g_1g_2}\right. \\& \left.+ \frac{{g_1}^2}{{Q_2}^2}   \left(   \frac{-3\eta}{(2\eta-1)\alpha_1}  - \frac{N_2  Q_1 k^{3/2}}{\eta}       \right)  -\frac{{g_1}^2}{{Q_2}^2} \left(\frac{5}{7} \frac{-3\eta}{(2\eta-1)\alpha_1}\right)\right] \\ & + \frac{4}{7}\sqrt{x_2} {(f^2(x_2))}^{5/2} \left[    \frac{-k^{5/2}{g_1}^2}{Q_1}  -    \frac{{g_2}^2}{Q_1 k^{1/2}} -  \frac{k}{Q_1} 2{g_1g_2}  +  \frac{k^{5/2}{g_1}^2}{Q_2}\right] \\&\left. +\frac{4}{7}\sqrt{x_1} {(f^1(x_1))}^{5/2} \left[- \frac{{g_1}^2}{Q_2}\right]\right].
\end{aligned}
 \end{equation}
Here we observe that the coefficient of $\frac{N_2}{\eta} \frac{df^2(x)}{dx}|_{x_2}$ vanishes; all the above eleven terms actually cancel. Also all the other seemingly difficult integrals boil down to a simple equation, which gives
\begin{equation}
\begin{aligned}
&U = -\frac{4}{\pi}\left( \frac {3\pi \eta}{2}\right)^{1/3} \frac {\frac{4}{3}g^2 \cdot \frac{4}{3}\alpha_s}{a}\times \\&  \bigg[ \left.-\frac{4}{7}\alpha_1 \frac {df^1(x)}{dx}\right|_{x_1} + \frac {4}{7} \sqrt{x_1}\left({f}^{{1}}(x_1)\right)^{5/2} g_1 \alpha_1  +\frac {4}{7} \sqrt{x_2}\left({f}^{{2}}(x_2)\right)^{5/2} g_2 \alpha_2  \bigg].
\end{aligned}
\end{equation}

So, the interesting thing we observe in both the kinetic and potential energy expressions is that there is no dependency on the derivative of TF function at the inner boundary, unlike the baryon case.


\subsection{$D$-meson family: Case 3}
The $D$-meson and multi quark-pair families of $D$-mesons consist of charm (or bottom) quarks (or antiquarks) and light antiquarks (or light quarks). All the quarks in this family can be represented as $Q\bar{q}$, $Q\bar{q}Q\bar{q}$, $Q\bar{q}Q\bar{q}Q\bar{q}$, and so on. We will study the systems where the heavier mass is a quark and the lighter mass is an antiquark in each member of the $D$-meson family. By the symmetry inherent in the TF equations, this also covers the antiparticle state where the particle and antiparticle are interchanged. There is asymmetry in total mass of quarks and antiquarks, unlike the other cases we have studied. Also, because quarks interact only with antiquarks, the mathematics is different than earlier cases. We have two different TF functions for the charm and anti-light quark, $f$ and $\bar{f}$, respectively. We assume there is a universal TF function $\bar{f}=k_0 f$ in the region where the TF functions overlap. This means they are related linearly in the smaller region where the TF function of the charmed quark is nonzero. Outside of this, only the TF function of the light antiquark exists.

For the region $0<x<x_2$, we have the differential equation
\begin{equation}
 \frac {d^2f(x)}{dx^2} = Q_0 \frac {\left(f(x)\right)^{3/2} }{\sqrt{x}},
\end {equation}
where
\begin{equation}
Q_0 = - \frac{3\eta \,g_0}{\left( 2\eta -1\right)\cdot k_0},
\end{equation}
and
\begin{equation}
k_0=\left(\frac{g_0\,\alpha}{\bar{g}_0\,\bar{\alpha}}\right)^{2/5},
\end {equation}
is the consistency condition. $\alpha$ is given by Eq.~(\ref{ratio}), and in this case $\bar{\alpha}=1$.

In region $x_2<x<x_1$ with two TF equations, one function is zero and the other becomes
\begin{equation}
\begin{aligned}
&\frac {d^2\bar{f}(x)}{dx^2} =  0, \\
\implies & \bar{f}(x)=c\cdot x +d ,
\end{aligned}
\end{equation}
where
\begin{equation}
c=k_0{f}'(x_2),
\end{equation}
and
\begin{equation}
d =k_0\left(f(x_2)-{f}'(x_2)\cdot x_2 \right) .
\end{equation}
Proceeding as before we obtained expressions for kinetic and potential energies. The energies are
\begin{equation}
\begin{aligned}
T= & \frac{12}{5\pi}\left( \frac {3\pi \eta}{2}\right)^{1/3} \frac {\frac{4}{3}g^2 \cdot \frac{4}{3}\alpha_s}{a}\times \\ & \Bigg[ \left. -\frac{10}{21}\alpha \frac {df(x)}{dx}\right|_{x_{2}} + \frac {8}{7} \sqrt{x_2}\left({f}(x_2)\right)^{5/2} g_0\alpha   +  \bar{g}_0\int_{x_2}^{x_1}{\frac{\left(\bar{f}(x)\right)^{5/2}}{\sqrt{x}}dx}  \Bigg],
\end{aligned}
\end {equation}
\begin{equation}
\begin{aligned}
U= &- \frac{4}{\pi}\left( \frac {3\pi \eta}{2}\right)^{1/3} \frac {\frac{4}{3}g^2 \cdot \frac{4}{3}\alpha_s}{a}\times \\ &\Bigg[\left.-\frac{4}{7}\alpha \frac {df(x)}{dx}\right|_{x_2}\frac{\bar{g}_0}{g_0} + \frac {4}{7} \sqrt{x_2}\left({f}(x_2)\right)^{5/2} \bar{g}_0\alpha +\frac{ \eta\, \bar{g}_0}{2\eta -1}   \int_{x_2}^{x_1}{\frac{\left(\bar{f}(x)\right)^{3/2}}{\sqrt{x}}dx}  \Bigg].
\end{aligned}
\end {equation}

As a consistency check, if we assume light and heavy quarks to have equal mass, we have found that the potential and kinetic energies with $\eta=1$ in Case 3 are equal to that of Case 1. Also, the non-degenerate Case 2 with $\eta=2$ is the same as degenerate Case 1 with $\eta=2$. We confirmed this both in our analytical and numerical results.

 \section{Method and Remarks}\label{sec7}

The phenomenological parameters we need for our model are the strong coupling constant $\alpha_s$, the bag constant $B$, the charm and bottom quark masses, $m_c$ and $m_b$, as well as the light quark mass, $m_1$. Previously, we used baryon phenomenology to obtain these parameters\cite{Baral}. In addition, we did a fitting using mesonic states that involves only the charm quarks\cite{singapore}. Here, we have included bottom quarks as well and obtained the fitted parameters. Since we do not yet include spin interactions in our model, we need to weight spin-split states to \lq\lq remove\rq\rq this interaction for our model fits.
 
Thus, we first fit the model expressions for the masses of charmonium states, bottomonium states, spin-weighted $D$-meson states, spin-weighted $B$-meson states, a spin-weighted combination of spin 1 likely tetraquark states involving charm and bottom quarks. In this way, we obtained the spin-weighted mass for these six states and then fit the parameters above to obtain the minimized chi-square. 
 
For the mass of charmonium and bottomonium we weighted masses of $1S$ states such that 
\begin{equation}
\frac{1}{4} \left(\eta_c\left(1S\right)\right) + \frac{3}{4}  \left(J/\Psi\left(1S\right)\right) = 3069 \text{ MeV}
\end{equation}
\begin{equation}
\frac{1}{4} \left(\eta_b\left(1S\right)\right) + \frac{3}{4}  \left(Y\left(1S\right)\right) = 9445 \text{ MeV}
\end{equation}
which we will refer to as Case 1-charm and Case 1-bottom mass.

To obtain Case 2-charm, we weighted spin 1 likely tetraquark states called the $Z_c(3900)$ and $X(4020)$. There are a number of charmonium-like exotic resonances which have been discovered in recent years\cite{PDGReview}. The ground state of this set is the $\chi_{c1}(3872)$, with a mass of 3871.7 MeV, discovered by the Belle Collaboration in 2003\cite{Belle0}. However, this state appears to be a isospin singlet, although the charge states might not have been seen yet\cite{Maiani}. On the other hand, the $Z_c(3900)$ resonance, with mass 3886.6 MeV, spin 1 and $C=-1$, is a triplet as we would expect from the hidden charm nonrelativistic tetraquark model. In addition, the fact that the $Z_c(3900)$ is above the $D^0\bar{D}^{*+}$ threshold makes it more likely that it is a true tetraquark rather than a molecular state of these same two particles. In the same way, the $X(4020)$ with mass 4024.1 MeV is a spin 1, $C=-1$ state which in this case is just above $D^{*0}\bar{D}^{*+}$ threshold. We will adopt these two particles as our spin-split charmed tetraquark states. Note that there are hidden bottom analogs of the $Z_c(3900)$ and $X(4020)$ states in the $Z_b(10610)$ and $Z_b(10650)$ states. We will use these to define the Case 2 $Z$-mesons. 

Note that a very simple model which can accommodate the $\chi_{c1}(3872)$, $Z_c(3900)$ and $X(4020)$ states has a spin interaction Hamiltonian given by
\begin{equation}
\begin{aligned}
H_{\text{spin}}= \kappa_1 (s_1(s_1+1)-3/2) &+ \kappa_2 (s_2(s_2+1)-3/2) + \kappa_3 (J(J+1)\\&
-s_1(s_1+1) -s_2(s_2+1)),
\end{aligned}
\end{equation}
where $s_1$ is the light quark spin, $s_2$ is the heavy quark spin and $J$ is the total spin. The three terms represent the light-light, heavy-heavy and light-heavy spin interactions, respectively. We would expect that $\kappa_2<\kappa_1, \kappa_3$ based on the quark masses. By neglecting $\kappa_2$ and fitting the three states mentioned, one predicts that there are additional hidden charm spin 0 states at $3720$ and $3887$ MeV, as well as a spin 2 state at $4176$ MeV, all with $C=1$. There is a very broad spin 0 candidate state called the $\chi_{c0}$ at $3862$ MeV as well as a narrower spin 0/2 sate called the $X(3915)$ at 3918 MeV, which are both apparently isospin singlets, but the other predicted states do not presently have candidates\cite{PDGReview}.

The Case 2 calculations are now simply,
\begin{equation}
\frac{1}{4}\times Z_c(3900) +\frac{3}{4}\times X(4020) = 3990 \text{ MeV},
\end{equation}
and
 \begin{equation}
 \frac{1}{4}\times Z_b(10610) +\frac{3}{4}\times Z_b(10650) = 10641 \text{ MeV},
 \end {equation}
which we will refer to as Case 2-charm and Case 2-bottom mass respectively. 

For  the mass of $D$ mesons we weighted the spin 0 charge states such that
\begin{equation}
D\left( \text{spin zero average} \right)=\frac{1}{2} \left(D^{+}\right) + \frac{1}{2}  \left(D^{0}\right) = 1866.5 \text{ MeV},
\end{equation}
and similarly for spin 1 $D^*$ mesons. We thus obtained the Case 3-charm mass
\begin{equation}
\frac{1}{4}D\left( \text{spin zero average} \right) + \frac{3}{4}D^*\left( \text{spin one average} \right) = 1973 \text{ MeV}.
\end{equation}
Similarly for B mesons, we obtained the Case 3-bottom value as 5313 MeV.

Using {\it Mathematica}, we fitted parameters such that the mass chisquare was minimized using a grid search, obtaining a total $\sqrt{\chi^2}= 86.1$ MeV spread over 6 masses. We solved the differential equations using an iterative implementation of {\bf NDSolve} in {\it Mathematica}. We obtained $\alpha_s = 0.346$, $B^{1/4} =107.6$ MeV, charm quark mass $m_c=1553$ MeV, and bottom quark mass $m_b=4862$ MeV. Our light quark mass, $m_1=306$ MeV, we take from our previous TF baryon fit\cite{WW2}. Our baryon paper found $\alpha_s = 0.371$ and $B^{1/4} =74.5$ MeV. Note that it would be premature to compare these sets of parameters as the baryon fit included spin interactions, which are quite significant, and the present fit does not. We will have additional comments on the inclusion of spin for TF mesons at the end of Section~\ref{sec8}. Table \ref{Table2554} shows the difference between expected and obtained masses.

\begin{table}[ht!]
\centering
 \caption{Comparison between masses used for fitting and those obtained after fitting.}
\begin{tabular}{ c c c } 
\\
\hline
 Fitted particle & Masses after fit(MeV)  & Masses used for fit(MeV) \\ 
\hline
 Case 1-charm &  $3049$ & $3069$  \\
Case 2-charm &   $4015$ & $3990$  \\
Case 3-charm &   $1960$ & $1973$  \\
Case 1-bottom &   $9469$ & $9445$  \\
Case 2-bottom &   $10653$ & $10641$  \\
 Case 3-bottom  &  $5239$ & $5313$  \\
 \hline
 
 \end{tabular}
\label{Table2554}
\end{table}

We will examine TF functions and energies for the three cases defined above. In nuclear physics, one examines the binding energy per nucleon in order to assess the stability of a given nucleus. We will do a similar investigation here. Thus, the important figure of merit in these evaluations is the total energy per quark, for if this increases as one adds more quarks, the family is unstable under decay to lower family members, whereas if it decreases, the family is stable. The static mass quark dependence of the the mesons does not play a role in these considerations and so will be left off.

\section{Results and Discussions}\label{sec8}

\begin{figure}[!htpb]
\centering
\includegraphics[scale=0.5,width=.85\textwidth]{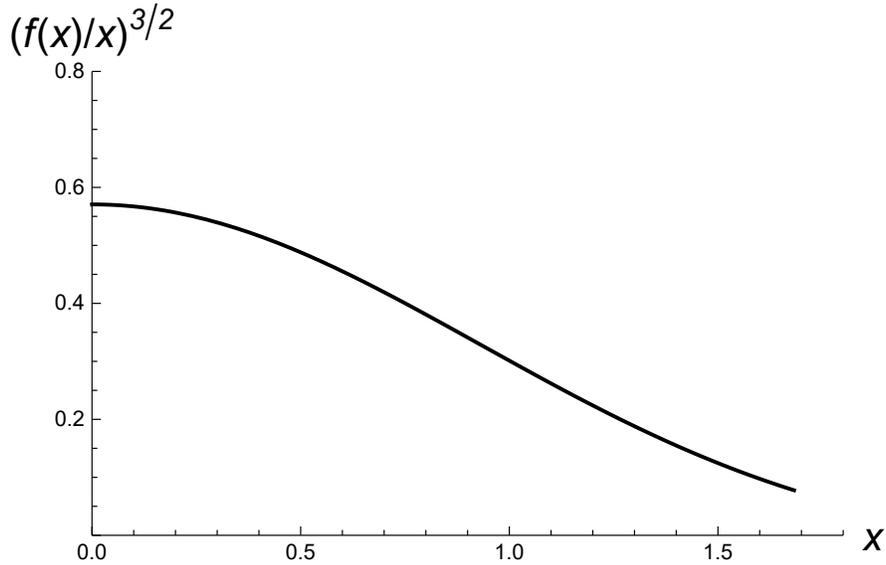}
\caption{TF density function of charmonium.}
\label{den1}
\end{figure}

\begin{figure}[!htpb]
\centering
\includegraphics[scale=0.5,width=.85\textwidth]{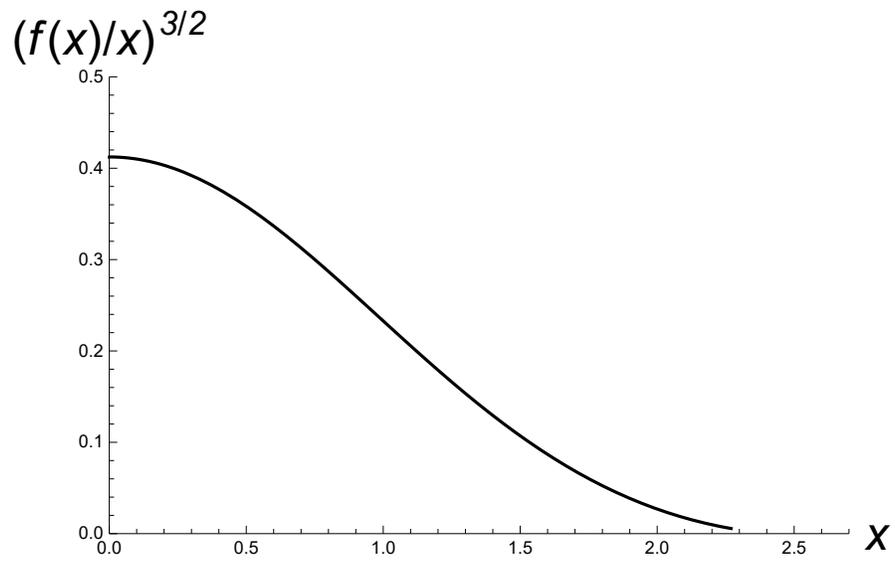}
\caption{TF density function of bottomonium.}
\label{den1b}
\end{figure}
First, let us first discuss the behavior of the TF density functions. The particle density function is proportional to $(f(x)/x)^{3/2}$ for all states. The charmonium and bottomonium function decreases smoothly in distance and has a discontinuity at the boundary as seen in Fig.~\ref{den1} and Fig.~\ref{den1b}, where the dimensionless $x$ variable is used. The discontinuity for the bottomonium density function actually is almost zero at the boundary.
\begin{figure}[!htpb]
\centering
\includegraphics[scale=0.5,width=.85\textwidth]{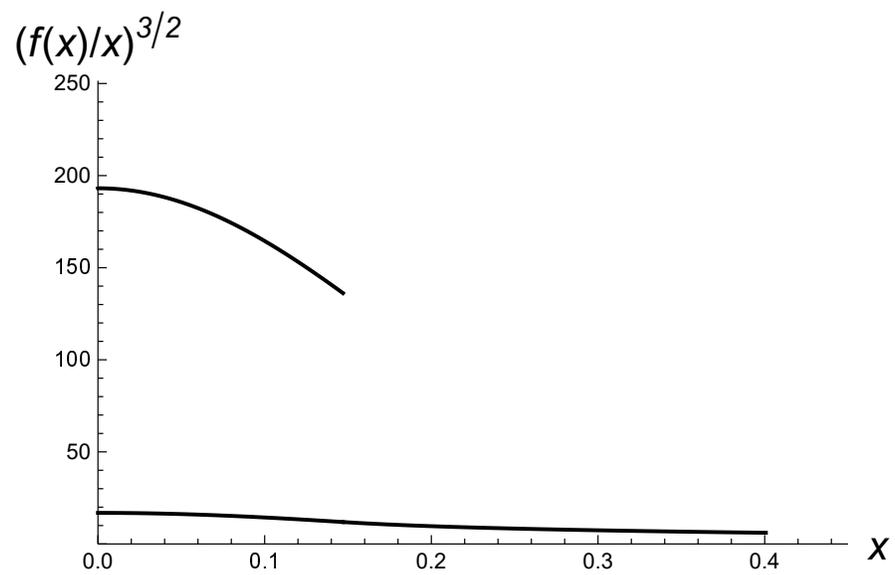}
\caption{TF density function of Case 2 mesons with charm quarks ($c\bar{c}u\bar{u}$).}
\label{den2}
\end{figure}

\begin{figure}[!htpb]
\centering
\includegraphics[scale=0.5,width=.85\textwidth]{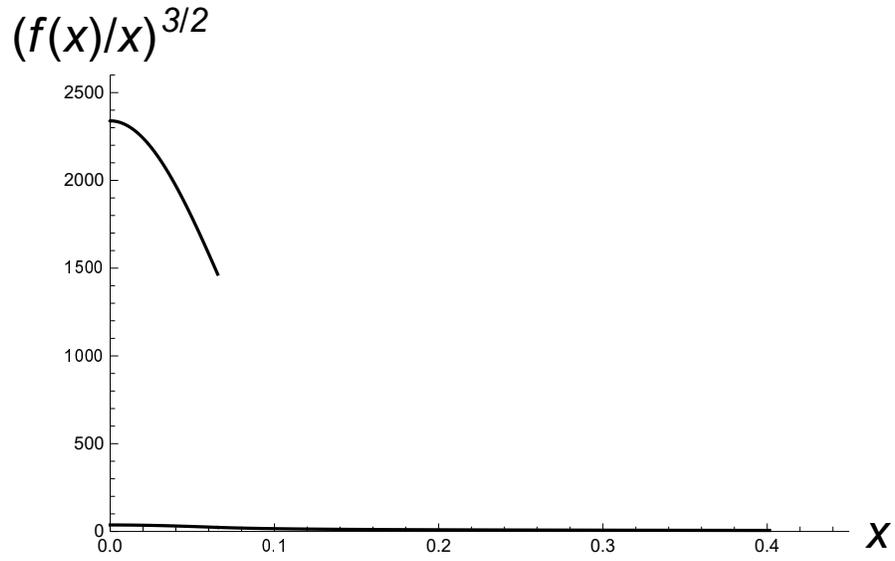}
\caption{TF density function of Case 2 mesons with bottom quarks ($b\bar{b}u\bar{u}$).}
\label{den2b}
\end{figure}

\begin{figure}[!htpb]
\centering
\includegraphics[scale=0.5,width=.85\textwidth]{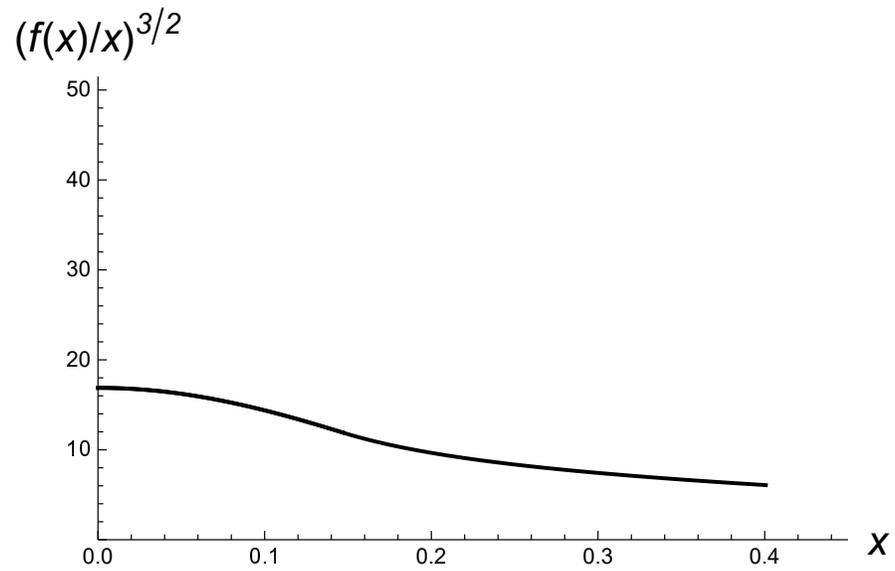}
\caption{TF density function of light quarks for the Case 2 mesons involving charm quarks ($c\bar{c}u\bar{u}$).}
\label{den12}
\end{figure}

\begin{figure}[!htpb]
\centering
\includegraphics[scale=0.5,width=.85\textwidth]{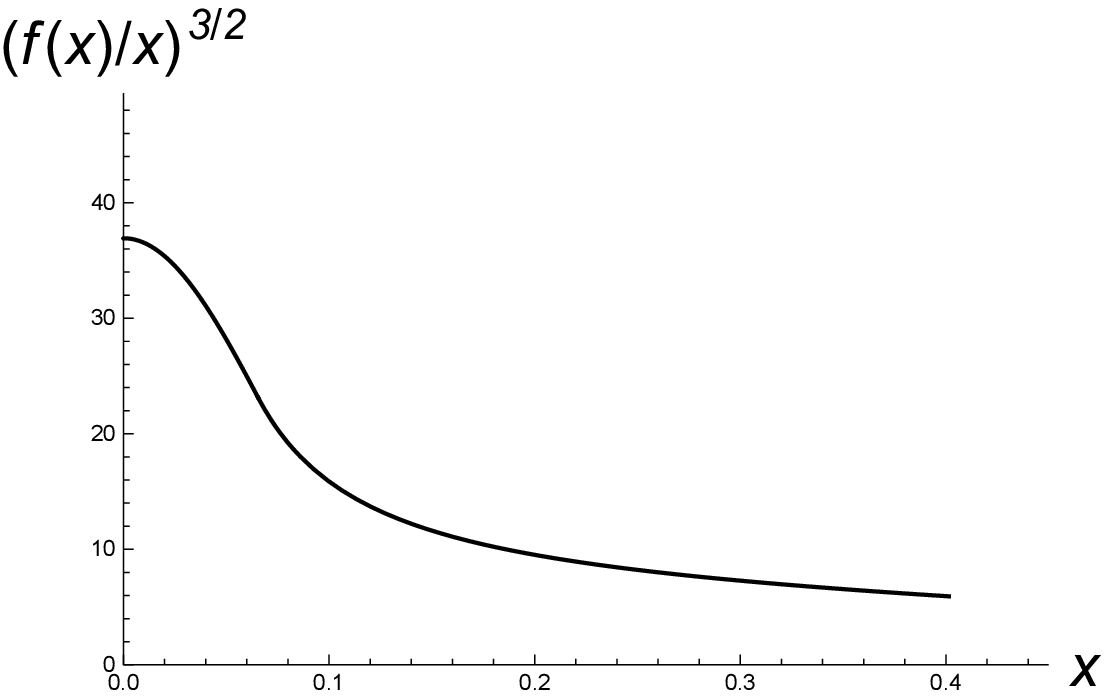}
\caption{TF density function of light quarks for the Case 2 mesons involving bottom quarks ($b\bar{b}u\bar{u}$).}
\label{den12b}
\end{figure}

\begin{figure}[!htpb]
\centering
\includegraphics[scale=0.5,width=.85\textwidth]{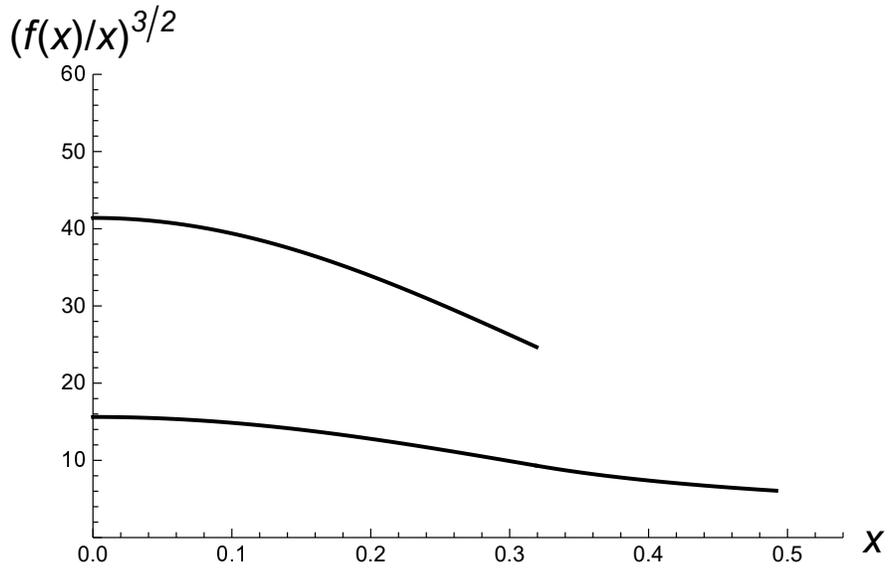}
\caption{TF density functions for the $D$-meson ($c\bar{u}$).}
\label{den3}
\end{figure}

\begin{figure}[!htpb]
\centering
\includegraphics[scale=0.5,width=.85\textwidth]{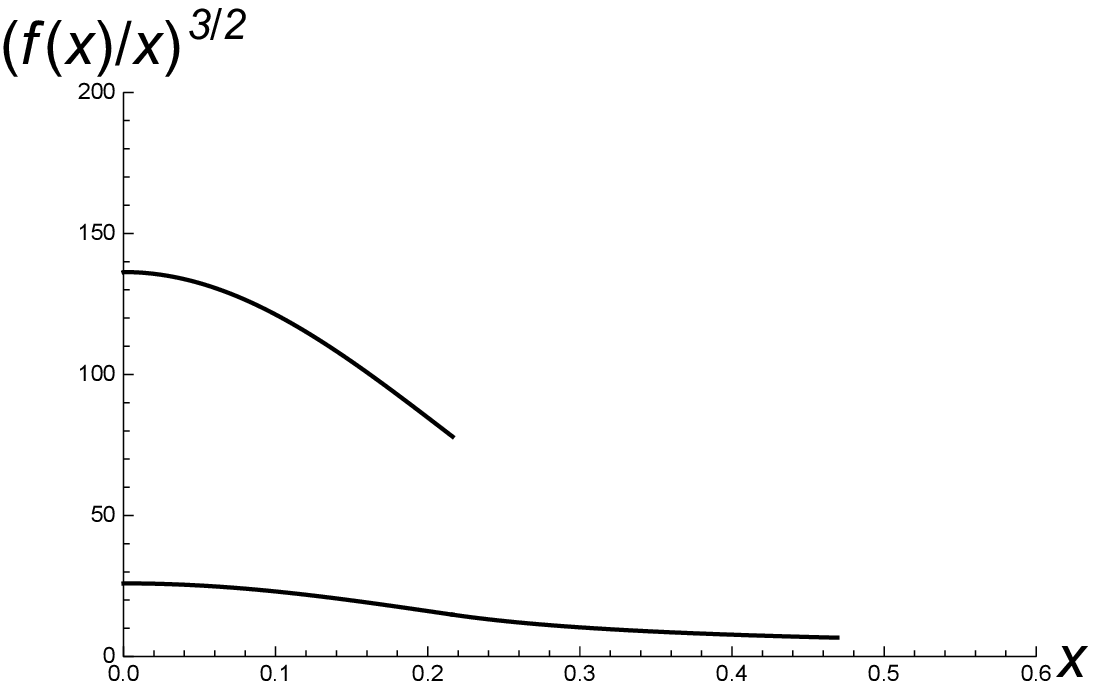}
\caption{Density of TF functions for the B-meson ($b\bar{u}$).}
\label{den3b}
\end{figure}

The density function of $Z$-mesons has a long tail for the light quarks, while for charmed quarks the value is large and is concentrated near the origin, as seen in Fig.~\ref{den2} and Fig.~\ref{den2b}. This suggests an atomic-like structure with heavy charm, anti-charm quarks at the center while light quarks and antiquarks spread out like electrons. Figs.~\ref{den12} and  Fig.~\ref{den12b} are an enlargement of the density function of the light quark TF function for the $Z$-meson. It drops down abruptly until it reaches the boundary of the heavy quark TF function, then inflects and decreases. In the case of D and B-mesons, Fig.~\ref{den3} and Fig.~\ref{den3b} respectively, the density function of light and heavy quarks are relatively closer. We can see that when charm is interchanged with bottom quarks the density function is more concentrated near the center, as one would expect.

\begin{figure}[!htpb]
\centering
\includegraphics[width=.85\textwidth]{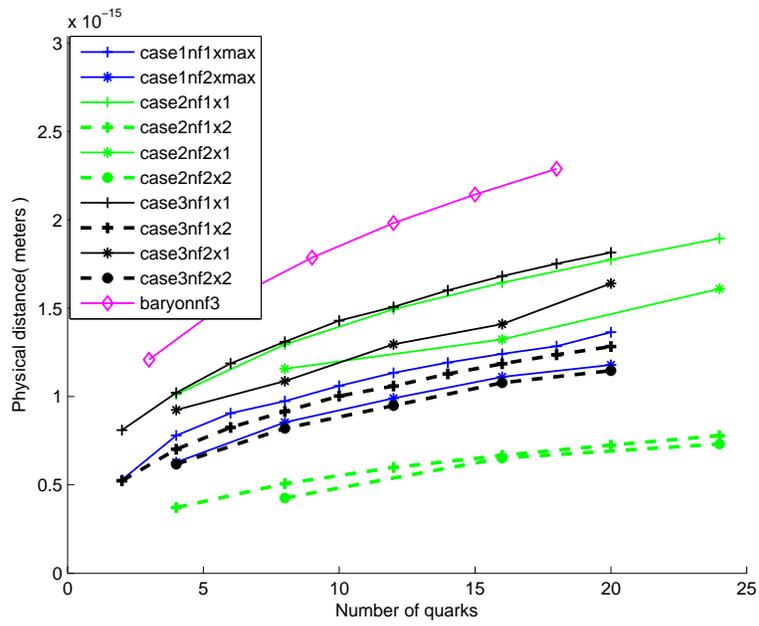}
\caption{Physical distance versus quark content, charm and light quark/antiquarks.}
\label{rad1}
\end{figure}

\begin{figure}[!htpb]
\centering
\includegraphics[width=.9\textwidth]{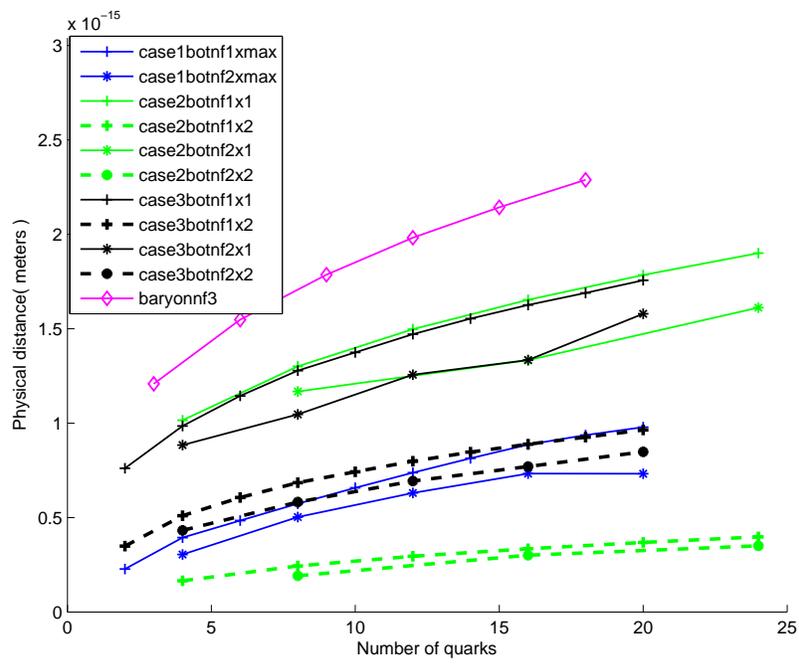}
\caption{Physical distance versus quark content, bottom and light quark/antiquarks.}
\label{rad1b}
\end{figure}

\begin{figure}[!htpb]
\centering
\includegraphics[width=.85\textwidth]{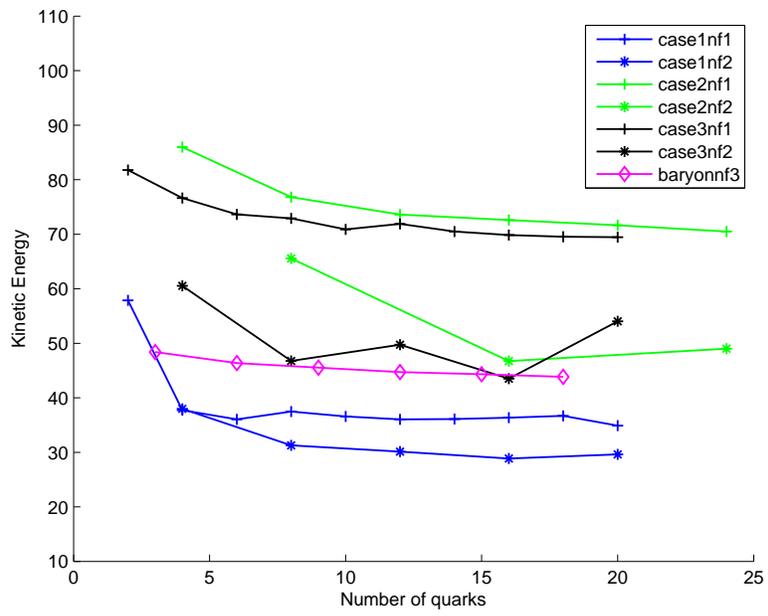}
\caption{Kinetic energy per quark versus quark content, charm and light quark/antiquarks.}
\label{kinetic}
\end{figure}

\begin{figure}[!htpb]
\centering
\includegraphics[width=.85\textwidth]{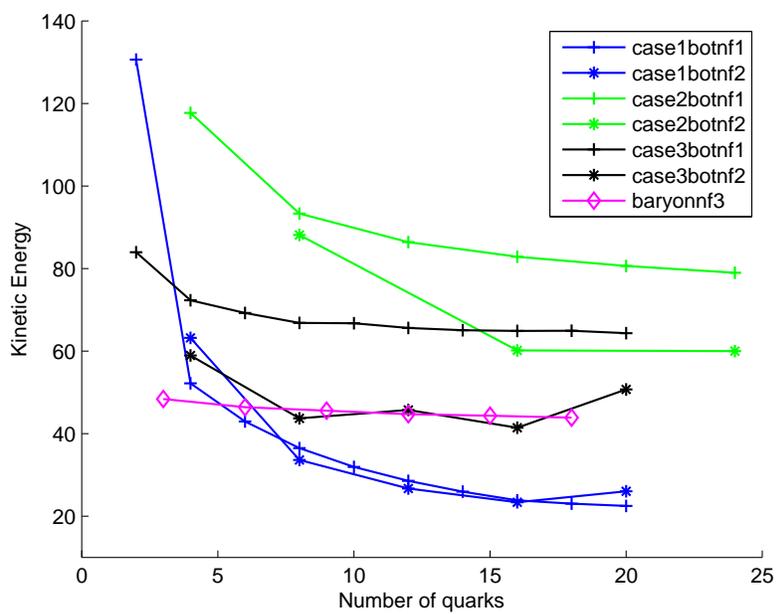}
\caption{Kinetic energy per quark versus quark content, bottom and light quark/antiquarks.}
\label{kineticb}
\end{figure}

\begin{figure}[!htpb]
\centering
\includegraphics[width=.85\textwidth]{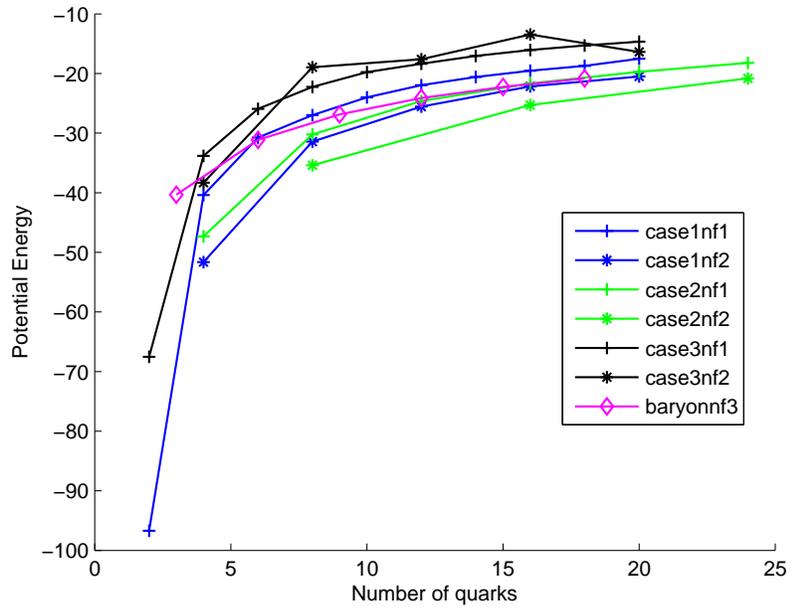}
\caption{Potential energy per quark versus quark content for charm and light quark/antiquarks.}
\label{potential}
\end{figure}

\begin{figure}[!htpb]
\centering
\includegraphics[width=.85\textwidth]{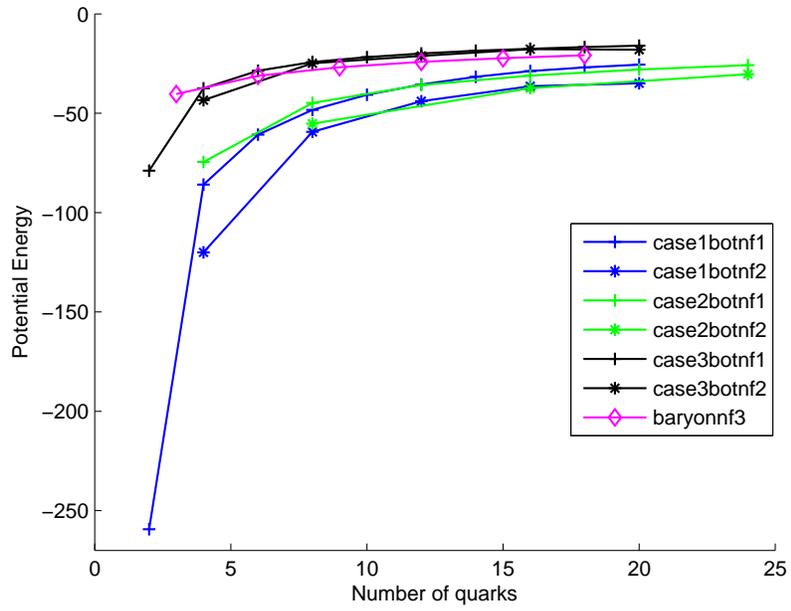}
\caption{Potential energy per quark versus quark content for bottom and light quark/antiquarks.}
\label{potentialb}
\end{figure}

\begin{figure}[!htpb]
\centering
\includegraphics[width=.85\textwidth]{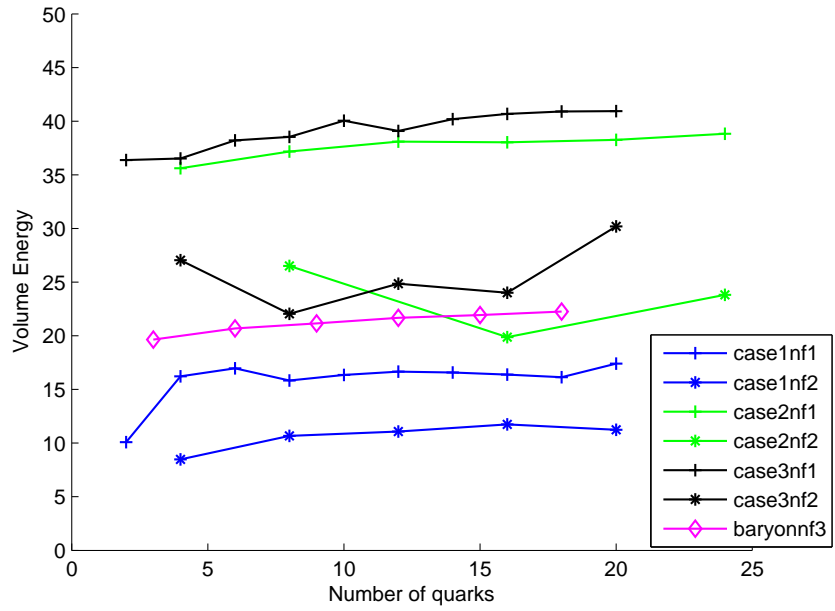}
\caption{Volume energy per quark versus quark content for charm and light quark/antiquarks.}
\label{volumepic}
\end{figure}

\begin{figure}[!htpb]
\centering
\includegraphics[width=.75\textwidth]{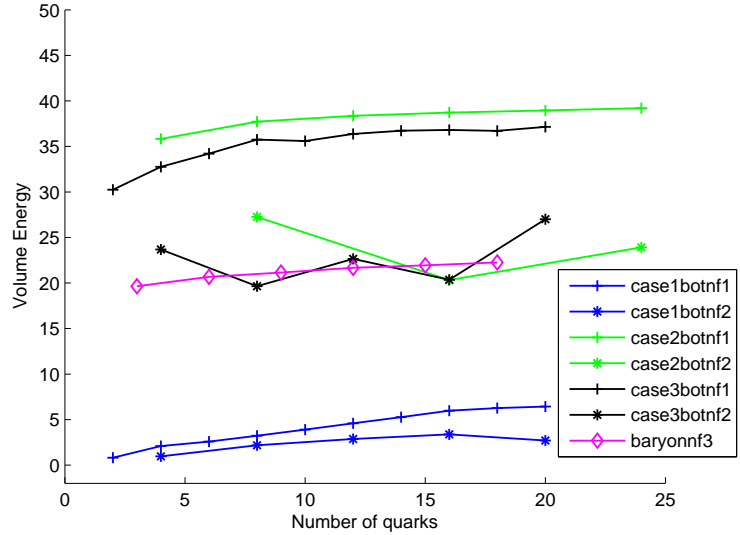}
\caption{Volume energy per quark versus quark content for bottom and light quark/antiquarks.}
\label{volumepicb}
\end{figure}

\begin{figure}[!htpb]
\centering
\includegraphics[scale=0.5,width=.75\textwidth]{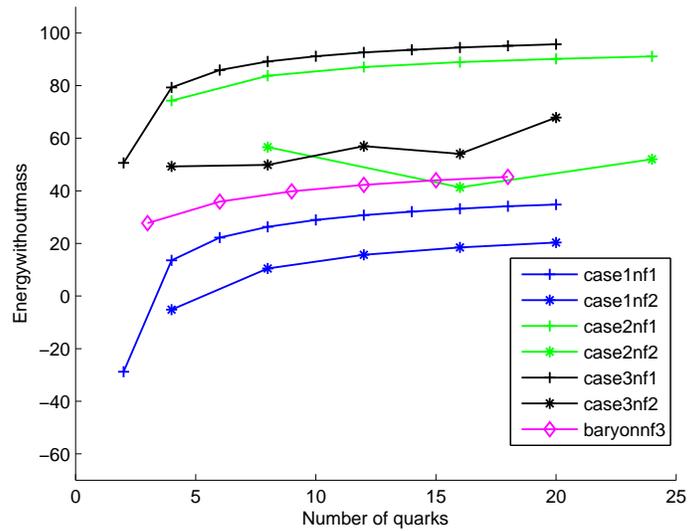}
\caption{Total energy per quark without mass term versus quark number  for charm and light quark/antiquarks.}
\label{energypic}
\end{figure}

\begin{figure}[!htpb]
\centering
\includegraphics[scale=0.5,width=.75\textwidth]{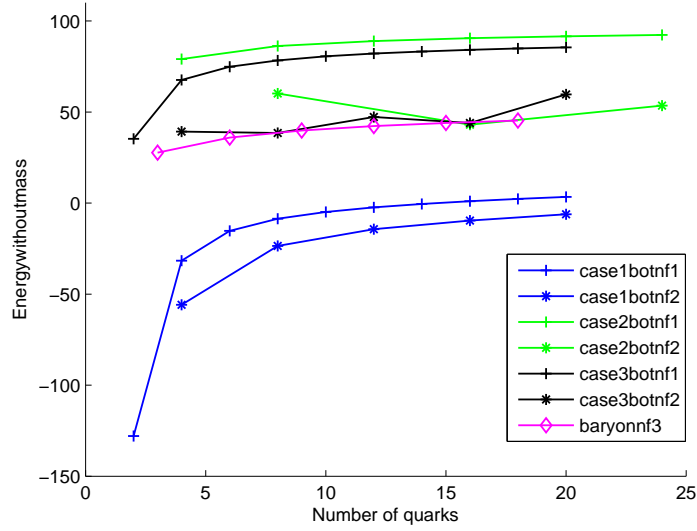}
\caption{Total energy per quark without mass term vs.~quark number for bottom and light quark/antiquarks.}
\label{energypicb}
\end{figure}
\begin{figure}[!htpb]
\centering
\includegraphics[scale=0.5,width=.75\textwidth]{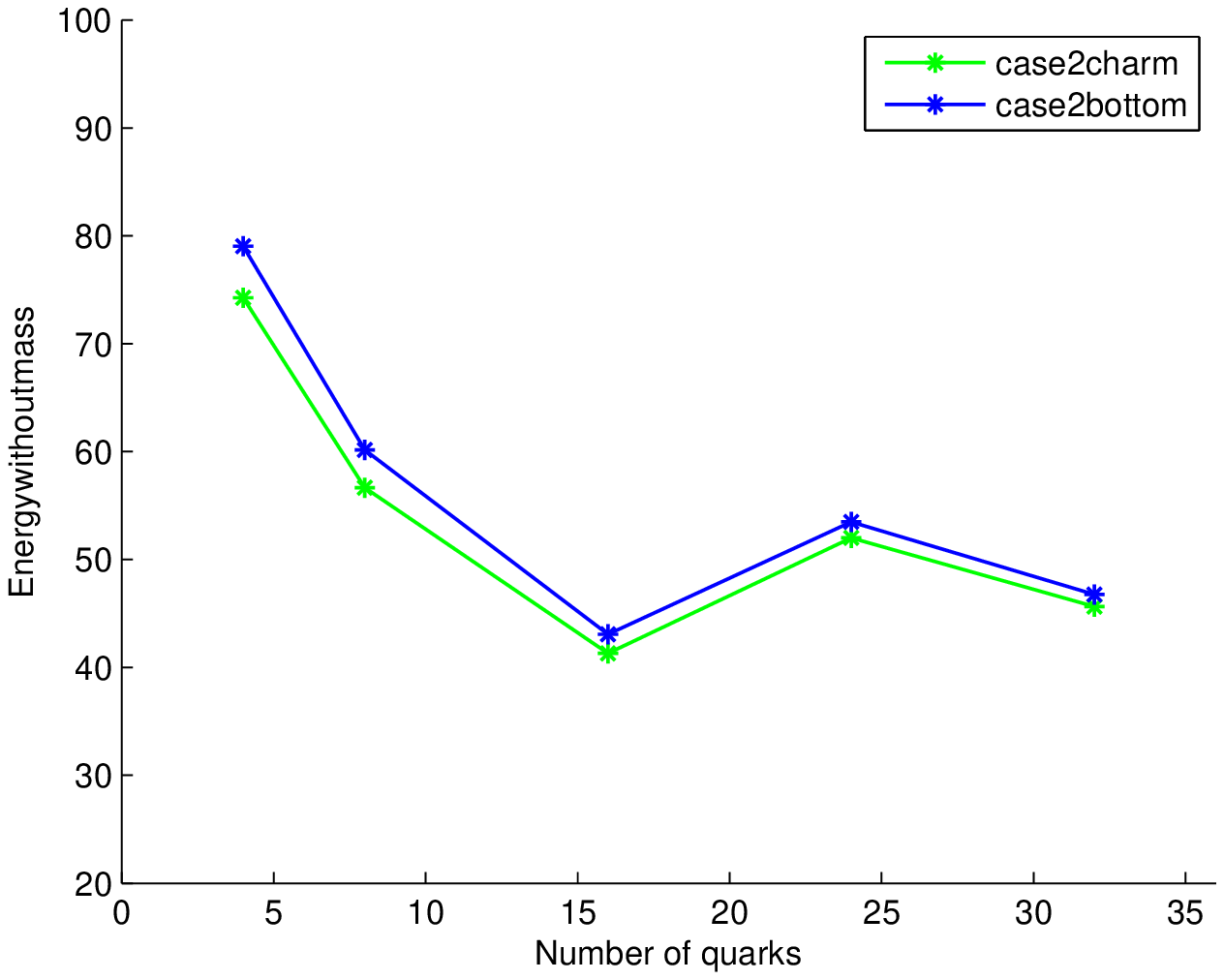}
\caption{Comparison of charm-light versus bottom-light quarks for Case 2. The results at $N=4$ come from the fitted Case 2-charm and -bottom states from Table~\ref{Table2554}.}
\label{charmbaryon2}
\end{figure}
\begin{figure}[!htpb]
\centering
\includegraphics[scale=0.5,width=.75\textwidth]{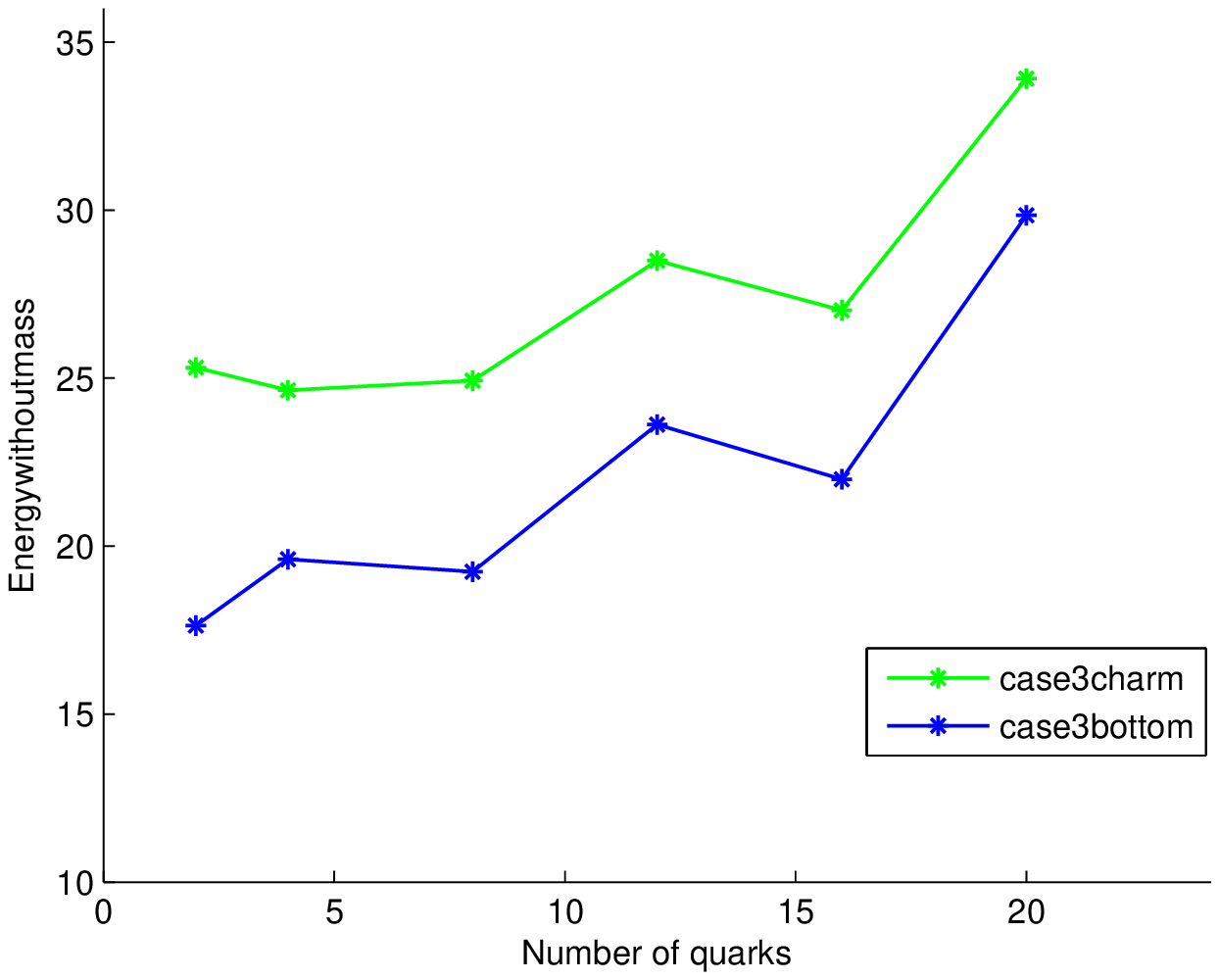}
\caption{Comparison of charm-light versus bottom-light quarks for Case 3. The results at $N=2$ come from the fitted Case 3-charm and -bottom states from Table~\ref{Table2554}.}
\label{charmbaryon3}
\end{figure}

We increased the quark content and compared density functions of a family of multi-mesons in all three cases. We observed similar density functions for a given multi-meson family regardless of the quark content. This is the embodiment of the Thomas-Fermi approach.

The Isgur-Wise symmetry methodology\cite{Isgur:1989vq} in $D$ and $B$ mesons is a capable tool in particle physics and has been implemented in various studies\cite{Roy:2012ng,Roy:2012dj,deRafael:1992tu,Ali:2017jda}. It arises when the heavy quark appears to the light quark degrees of freedom as a static color source\cite{Kenway:1993yp}. There is a natural question of how close our simulations for multiquarks approach such a heavy-light symmetry scenario. As one can see by an examination of the TF functions in Figs.~\ref{den2b} and \ref{den3b}, there is never a case in which one can consider the radial extent of the heavy quark density function exceptionally compact, and thus kinematically nondynamical, compared to the light quark. Although Figs.~\ref{den3} and \ref{den3b} refer to $D$ and $B$ systems, remember that the TF approach actually embodies multiquarks, as discussed above. Also, note that the Case 2 and Case 3 heavy quark profiles are very different, reflecting the different interactions present among the quarks and antiquarks. The Case 2 heavy quarks interact with both heavy and light antiquarks, whereas the Case 3 heavy quarks interact only with the light antiquarks. Since the interactions are attractive this makes the Case 2 heavy quark profile more compact than Case 3. For light atomic elements one can neglect electron-electron interactions to lowest order, which is actually more like Case 3. Thus, even though the Case 2 heavy quark profile is more compact like an atomic nucleus, the interactions make the Case 2 mesons less atomic-like than Case 3. These TF functions and interactions suggest it is unlikely that a nondynamical heavy flavor symmetry will be manifest in heavy-light multiquark states for charm and bottom quarks.

Fig.~\ref{rad1} and Fig.~\ref{rad1b} includes 11 possibilities for the physical radius. The physical radius is plotted versus quark number and compared with a generic baryon with three degenerate light flavors. From Eq.~(\ref{fn2}) we can see that the radius is proportional to the product of $\eta^{2/3}$ and the dimensionless radius $x$. As the quark content increases, the dimensionless radius, $x$, becomes smaller, whereas $\eta^{2/3}$ increases. In most cases, the result is an increase in radius with increasing quark content. We also observe that the curve of the radius plot for each case tends to flatten out for larger numbers of quarks. {\bf case1nf1xmax} refers to multi quark-pair families of charmonium with no degeneracy. In this case all the charm quarks have the same spin and hence cannot occupy the same state. {\bf case1nf2xmax} is instead the plot of the charmonium family with a degeneracy of two. In this case, spin up and down are assigned to a pair of charm quarks, and thus they do not occupy the same state. Thus, the physical radius of {\bf case1nf1xmax} is larger than {\bf case1nf2xmax} reflecting this fact. Note that the dotted lines refer to the inner boundary associated with the charmed quark in Cases 2 and 3. For Case 2 the difference between dotted and continuous lines is the largest. That means, like the $Z$-meson, all the higher quark family members have heavy charm-anticharm concentrated at the center while light and anti-light quarks are spread throughout. {\bf case2nf1x2} refers to the radius plot of the inner boundary of the multi quark-pair family of $Z$-mesons with degeneracy of one while {\bf case2nf2x2} refers to the same plot with degeneracy of two, and similarly for Case 3. In all cases we see that the plot of physical radius is larger for degeneracy one compared to degeneracy two. The $Z$ and $D$-meson family members are found to have equally large outer boundaries. For a given quark number, the outer radius of all types of mesons was found to be smaller than the generic baryon. Note that these radii are determined by the size of the TF functions; the electromagnetic radii have not yet been evaluated.

There are 3 types of energies in this model: kinetic, potential and volume. The kinetic energy per quark (in MeV) depends strongly on the meson family, as seen in Fig.~\ref{kinetic} and Fig.~\ref{kineticb}. We see that the energy per quark is relatively small and tends to decrease slowly, which seems to provide some justification for this nonrelativistic model. Fig.~\ref{potential} and Fig.~\ref{potentialb} shows the corresponding graph for the potential energy. The energies rise quickly from rather large negative values to saturate near small negative values for all cases. Fig.~\ref{volumepic} and Fig.~\ref{volumepicb} shows the graph for volume energies. The graphs are similar when we interchange bottom with charm quarks. The {\bf case2nf2} and {\bf case3nf2} results in these figures deserve some extra comments. These lines are determined by the most degenerate state, and thus the smallest energy per quark, available for the given quark content. If we denote the total number of quarks and antiquarks as $N$ ($=2\eta$), the $N=$ 8, 16 and 24 quark cases for {\bf case2nf2}, which all have $g_2=2$, are associated with $g_1=2$, $g_1=4$ and $g_1=3$, respectively. Likewise, the $N=$ 4, 8, 12, 16 and 20 quark cases for {\bf case3nf2} are associated with $g_1=2$, $g_1=4$, $g_1=3$, $g_1=4$ and $g_1=2$, respectively.

Figs.~\ref{energypic} and \ref{energypicb} show the total energy per quark without the mass term, i.e., the sum of kinetic, potential and volume energy, plotted against the quark content. The generic baryon rises slowly and smoothly for increasing quark content, implying these are unstable; i.e., a higher quark content state can decay into lower members of the same family. The Case 1 mesons rise quickly, and then continue the rise more slowly; these are also unstable. The Case 2 and 3 nondegenerate mesons rise less quickly but still continue a slow increase for higher quark numbers. 

The lowest mass Case 2 and 3 patterns have been highlighted in Figs.~\ref{charmbaryon2} and \ref{charmbaryon3}. For Case 2 we can observe negative slope from $N=4$ to $N=8$  and again from $N=8$ to $N=16$ when we account for degeneracy. Here we have plotted five points for Case 2 with both charm-light quarks and bottom-light quarks. The charm and bottom curves are very close to one another. After that there is an inflection at $N=24$. This hints at the possible existence of stable multi-mesons with $N=8$ and also with $N=16$, which we term octaquarks and hexadecaquarks. On the other hand, the Case 3 charm and bottom patterns are separated from one another, but are still very similar. There is an overall upward trend to this data, although there is an isolated downward dip from $N=12$ to $N=16$. We will comment on the $N=4$ results from this figure below.

There are lattice QCD results for Case 3 tetraquarks. Ref.~\citenum{Nilmani} has two-point function results for quark flavor content $ud\bar{c}\bar{c}$ with spin 1 and $uu\bar{c}\bar{c}$ content for spin 0 for dynamical pion masses ranging from $257-688$ MeV. Their continuum extrapolations show binding of $-23.3\pm 11.4$ MeV for the spin 1 sector and antibinding (resonance energy) of $25.9\pm 10.9$ MeV above threshold in the spin 0 sector. On the other hand, the $ud\bar{b}\bar{b}$ with spin 1 has a reported binding of $-143.3\pm 33.9$ MeV and $uu\bar{b}\bar{b}$ has a small spin 0 binding of $-5.5\pm 17.7$ MeV. We can not compare directly with these spin sector results because we have not yet constructed the spin states as in our baryon study\cite{WW2}. The best we can do at this stage is to point out that our Case 3 results from Fig.~\ref{charmbaryon3} double charm quarks show a very small downward jump in energy per quark from $N=2$ to $N=4$ of $-0.67$ MeV, corresponding to a total binding energy of $-2.68$ MeV, whereas our double bottom results from the same figure show a small upward jump of $1.98$ MeV, leading to total antibinding of $7.9$ MeV. Our lowest energy states are always the most degenerate, which would correspond to the spin 0 tetraquark state in the doubly heavy sector. Although it is a bit premature to compare, the reported Monte Carlo error bars in this reference can accommodate our results at the $\sim 2\sigma$ (double charm case) or $\sim 1\sigma$ (double bottom case) level. More precise studies will be necessary to make definitive conclusions concerning the binding state of these types of particles.

In our future work, we will need to include the spin interactions. In mesonic models, the color probabilities only allow quark-antiquark interactions, which greatly simplifies the spin calculations. A derivation of the color magnetic interaction similar to that in Ref.~\cite{WW2} gives
\begin{equation}
E_m=\frac{8\pi}{3}\frac{9\times \frac{4}{3}g^2}{(2\eta -1)}\sum_{{\cal I}<{\cal J}}   \gamma_{\cal I} \bar{\gamma}_{\cal J}  \sum_{i=1}^{N_{\cal I}}\sum_{j=1}^{N_{\cal J}}   (S_z^{\cal I})_i(\bar{S}_z^{\cal J})_j  \int d^3r\, \left(   \frac{n^{\cal I}_i(r)\bar{n}^{\cal J}_j(r)}{N_{\cal I}\bar{N}_{\cal J}}  \right) . \label{spinenergy}
\end{equation}
The sums are over generalized flavor indices $\cal I, J$ (which include spin) as well as particle indices $i,j$. The $(S_z^{\cal I})_i, (\bar{S}_z^{\cal J})_j$ are the spin components and the $\gamma_{\cal I},\bar{\gamma}_{\cal J}$ factors in \ref{spinenergy} above are given by
\begin{equation}
\gamma_{\cal I}=\frac{\text{g}}{2m_{\cal I}c}, \bar{\gamma}_{\cal J}=\frac{\text{g}}{2\bar{m}_{\cal J}c},
\end{equation}
where $m_{\cal I},\bar{m}_{\cal J}$ is the quark or antiquark mass and $\text{g}$ is the color gyromagnetic factor. Note that the energy shift in \ref{spinenergy} is for each TF configuration; as in \cite{WW2} the particle states are weighted sums of the various TF configurations. As pointed out above, including the spin interaction will affect the other model parameters and energy extrapolations. However, comparison of the previous baryon, Ref.~\cite{WW1}, with newly fit meson parameters will allow us to evaluate systematic uncertainties within the TF quark model approach in future spectroscopic work.

\section{Conclusions and Acknowledgements}\label{sec9}

We have initiated the study of multi quark-pair mesons using the TF quark model. After specifying the explicit interactions and summing on colors, we have formulated system interactions and energies in a mean field approximation. We have investigated three cases of mesonic states: charmonium family, $Z$-meson family and $D$-meson family, as well as their bottom quark analogs. We have not yet included explicit spin interactions in our model, but we can take one level of degeneracy into account in our two-TF function construction. 

We have observed interesting patterns of single quark energies. Similar to our findings for baryons, the energy per quark is slowly rising for the nondegenerate Case 1 and 3 mesons, implying family instability. Our Case 3 finding for tetraquarks and other most degenerate states can not be considered definitive because of the lack of spin splitting terms in our interaction, but the overall trend argues against a family of such higher quark number states. Our Case 2 findings are the most interesting of our study. It is the only case where we see an actual sustained {\it decrease} in the energy of introduced quark pairs. This would indicate that stable octaquark and hexadecaquark versions of the charmed and bottom $Z$-meson exist. However, the energy trend is not smooth and the spin interactions have also not been taken into account. These findings are complicated dynamical results and could not have been predicted from first principles.

Our first order of business as we extend the model will be the inclusion of explicit spin interactions, to bring our meson model to the same level of development as the baryon model. These interactions can be determined from the nonrelativistic ground state wave functions of these states and the use of Eq.~(\ref{spinenergy}). A further extension of this model would be to examine mixed baryonic-mesonic states such as pentaquark families.

We thank the Baylor Quantum Optics Initiative and the University Research Committee of Baylor University for their partial support of this project. We also gratefully acknowledge discussions with N.~Mathur as well as helpful considerations from G.~Chandra Kaphle.

\newpage
\bibliography{mybibfile}{}

\end{document}